\documentclass[a4paper,fleqn,usenatbib]{mnras}

\usepackage{newtxtext,newtxmath}

\usepackage[T1]{fontenc}
\usepackage{ae,aecompl}

\usepackage{graphicx}	
\usepackage{amsmath}	
\usepackage{amssymb}	
\usepackage[ugly]{nicefrac}
\usepackage{pdfpages}
\usepackage{longtable}

\newcommand{\angstrom}{\textup{\AA}}

\title[Ionised gas outflows in MaNGA AGN]{Ionised gas outflow signatures in SDSS-IV MaNGA active galactic nuclei}

\author[Dominika Wylezalek]{Dominika Wylezalek$^{1,2}$\thanks{E-mail: dwylezal@eso.org},
Anthony M. Flores$^{2}$,
Nadia L. Zakamska$^{2}$,
Jenny E. Greene$^{3}$, \newauthor
Rogemar A. Riffel$^{4,5}$
\\
$^{1}$European Southern Observatory, Karl-Schwarzschildstr. 2, D-85748 Garching bei Mu\"{u}nchen, Germany\\
$^{2}$Department of Physics \& Astronomy, Johns Hopkins University, Bloomberg Center, 3400 N. Charles St., Baltimore, MD 21218, USA \\
$^{3}$Department of Astrophysical Sciences, Princeton University, Princeton, NJ 08544, USA\\
$^{4}$Departamento de F\'{i}sica, CCNE, Universidade Federal de Santa Maria, Av. Roraima, 1000, 97105-900 Santa Maria, RS, Brazil \\
$^{5}$Laborat\'{o}rio Interinstitucional de e-Astronomia - LIneA, Rua Gal. Jos\'{e} Cristino 77, 20921-400 Rio de Janeiro, RJ, Brazil
}
\date{Accepted XXX. Received YYY; in original form ZZZ}

\pubyear{2017}

\begin{document}
\label{firstpage}
\pagerange{\pageref{firstpage}--\pageref{lastpage}}
\maketitle

\begin{abstract}
The prevalence of outflow and feedback signatures in active galactic nuclei (AGN) is a major unresolved question which large integral field unit (IFU) surveys now allow to address. In this paper, we present a kinematic analysis of the ionised gas in 2778 galaxies at $z \sim 0.05$ observed by SDSS-IV MaNGA. Specifically, we measure the kinematics of the [OIII] $\lambda$5007\AA\ emission line in each spatial element and fit multiple Gaussian components to account for possible non-gravitational motions of gas. Comparing the kinematics of the ionised gas between 308 MaNGA-selected AGN that have been previously identified through emission line diagnostics and sources not classified as AGN, we find that while 25\% of MaNGA-selected AGN show [OIII] components with emission line widths of $> 500$~km~s$^{-1}$ in more than 10\% of their spaxels, only 7\% of MaNGA non-AGN show a similar signature. Even the AGN that do not show nuclear AGN photoionisation signatures and that were only identified as AGN based on their larger scale photoionisation signatures show similar kinematic characteristics. In addition to obscuration, another possibility is that outflow and mechanical feedback signatures are longer lived than the AGN itself. Our measurements demonstrate that high velocity gas is more prevalent in AGN compared to non-AGN and that outflow and feedback signatures in low-luminosity, low-redshift AGN may so far have been underestimated. We show that higher luminosity MaNGA-selected AGN are able to drive larger scale outflows than lower luminosity AGN. But estimates of the kinetic coupling efficiencies are $\ll 1$\% and suggest that the feedback signatures probed in this paper are unlikely to have a significant impact on the AGN host galaxies. However, continuous energy injection may still heat a fraction of the cool gas and delay or suppress star formation in individual galaxies even when the AGN is weak. 

\end{abstract}




\section{Introduction}

The energy output of actively accreting supermassive black holes (active galactic nuclei, AGN) has become a critical ingredient in modern galaxy formation theories. Powerful AGN (quasars, $L_{bol} > 10^{45}$~erg/s) can heat and photo-ionise gas tens of kiloparsecs away, and even well into the circum-galactic medium \citep{john15, rudie17} in a process known as radiative feedback. Furthermore, radiatively driven nuclear winds \citep{murr95} or jets can launch galaxy-wide outflows. Such mechanical feedback processes can aide in establishing~the black hole vs.\ bulge correlations, can effectively quench star formation activity, and~--~most importantly -- set the upper limit to the masses of galaxies \citep[e.g.]{ferr00, crot06, fabi12, korm13}. However, constraining the power and reach of such feedback processes exerted by black holes onto their hosts remains a major unresolved issue in modern extragalactic astrophysics.

The critical role of quasars in galaxy formation was hypothesised two decades ago \citep{silk98}, yet this paradigm only recently obtained observational support, much of it on the basis of IFU observations \citep{rupk11,liu13b,harr14,carn15}. There is increasing evidence that in powerful AGN the main interaction with the gas is through winds, which are inhomogeneous, complex multi-phase phenomena, with different gas phases observable in different spectral domains \citep{heck90, veil05}. Most of our current knowledge about AGN-driven outflows comes from mapping the kinematics of the warm ionized gas phase via optical emission lines such as [O\,III]~5007~\AA. The signature of galaxy-wide winds is that of gas on galactic scales moves with velocities inconsistent with a dynamical equilibrium with the host galaxy or disk rotation \citep{rupk13,liu13b, wyle16a, wyle18}. 

Integral field unit (IFU) surveys now offer new possibilities in characterising outflow signatures for statistically significant samples. The SDSS-IV \citep{Blanton_2017} survey Mapping Nearby Galaxies at APO \citep[MaNGA;][]{Bundy_2015, Drory_2015, Law_2015, Yan_2016a, Yan_2016b, Wake_2017} is a new optical fibre-bundle IFU survey and will obtain IFU observations of 10,000 galaxies at $z \lesssim 0.1$ over the next few years, allowing an extensive investigation of the spatial dimension of galaxy evolution. The goals of the survey are to improve the understanding on the processes involved in galaxy formation and evolution over time.

MaNGA also allows to take full advantage of the spatial dimension of AGN ionisation signatures \citep{penny18, rembold17, Sanchez18, wyle17a, wyle18}. \citet{wyle18} have recently developed spatially resolved techniques tailored to the MaNGA data for identifying signatures of AGN. Out of 2778 galaxies in the parent sample, they identify 303 AGN candidates which show signatures of gas ionised by relatively hard radiation fields inconsistent with star formation. While the authors show that $\sim 10$\% of low redshift galaxies currently host low- to intermediate-luminosity AGN based on photoionisation diagnostics, it remains unclear if and to what extent these AGN impact the gas kinematics through AGN-driven winds. Additionally, \cite{wyle18} show that about a third to half of the MaNGA-selected AGN candidates would not have been selected based on the SDSS-III single-fibre observations since AGN ionisation signatures are only prevalent beyond the 3 arcsec coverage of the single-fibre spectra. Reasons for such signatures can be manyfold (heavy circumnuclear obscuration, off-nuclear AGN, dominant nuclear SF signatures, relic AGN) and are currently under investigation. 

A particularly intriguing possibility is that some of the AGN candidates are relic AGN. In such objects the nuclear activity subsided some time ago, but the photo-ionisation signatures at large distances persist for $10^4-10^5$ years due to light-travel delays and radiative timescales of emitting gas \citep{lint09, Schawinski_2015, Sartori_2016, Keel_2017}. In relic AGN, kinematic signatures of previous AGN activity may be longer lived than nuclear photoionisation signatures \citep{Ishibashi15}. Investigating the ionised gas kinematics in currently active AGN with nuclear AGN signatures and relic AGN candidates is therefore of great interest with respect to outflow timescales and outflow propagation. 

In this paper, we investigate the prevalence of ionised gas outflow signatures in MaNGA-selected AGN. This work focuses on a detailed kinematic analysis of the [O\,III]$\lambda$4959,5007\AA~doublet. We first develop a spaxel-based fitting algorithm allowing broad secondary components in the emission line profile to be accounted for. Our goal is to improve on the kinematic measurements previously made by the survey pipeline and use them in conjunction with a sample of independently identified AGN candidates from \citet{wyle18}. 

The paper is organised as follows: Section 2 introduces the MaNGA survey and structure of the available data. Section 3 presents the spectroscopic fitting procedure, while Section 4 presents the kinematic analysis. In Section 5 we discuss the identification of ionised gas outflow signatures and their prevalence in MaNGA-selected AGN and non-AGN. In Section 6 we present our conclusions. To statistically compare distributions, we use the two-sample Kolmogorov-Smirnov test and report $p$, the probability of the null hypothesis that the two samples are drawn from the same distribution. Low $p$ values ($ p < 0.01$) mean that the two samples are statistically different. Throughout the paper we use $H_{0} = 72$~km~s$^{-1}$~Mpc$^{-1}$, $\Omega_m=0.3$,$\Omega_{\Lambda}=0.7$.

\section{Data}

\subsection{The MaNGA Survey and Data Products}

Mapping Nearby Galaxies at Apache Point Observatory (MaNGA) is a spectroscopic survey as part of the Sloan Digital Sky Survey-IV (SDSS-IV). MaNGA is a two-dimensional spectroscopic survey that uses Integral Field Unit (IFU) observations to take multiple spectral observations of each galaxy in the 3,600 -- 10,000\angstrom\ range using the BOSS Spectrograph \citep{Gunn_2006, smee13} at $R \sim 2000$. Fibers are arranged into hexagonal groups, with bundle sizes ranging from 19 -- 127 fibres, depending on the apparent size of the target galaxy (which corresponds to diameters ranging between 12\arcsec\ to 32\arcsec), leading to an average footprint of $400-500$~arcsec$^{2}$ per IFU. The fibres have a size of 2\arcsec\ aperture (2.5\arcsec\ separation between fibre centres), which at $z\sim 0.05$ corresponds to $\sim 2$ kpc, although with dithering the effective sampling improves to $1.4$\arcsec. The current data release DR14 \citep{dr14_2017} contains 2778 galaxies at $0.01 < z < 0.15$ with a mean $z \sim 0.05$. Over the next three years, MaNGA will have obtained observations of $\sim$10,000 galaxies at $z\la 0.15$ and with stellar masses $>10^9 M_{\odot}$.

The MaNGA Data Reduction Pipeline (DRP) produces sky-subtracted spectrophotometrically calibrated spectra and rectified three-dimensional data cubes that combine individual dithered observations \citep[for details on MaNGA data reduction see][]{Law_2016} with a spatial pixel scale of 0.5 \arcsec\ pixel$^{-1}$. The median spatial resolution of the MaNGA data is 2.54 \arcsec\ FWHM while the median spectral resolution is $\sim 72$ km/s \citep{Law_2016}. The MaNGA Data Analysis Pipeline \citep[DAP, ][]{Yan_2016a, Westfall_2017} is a project-led software package used to analyse the data products provided by the MaNGA DRP, providing the collaboration and public with survey-level quantities, such as stellar-population parameters, kinematics and emission-line properties for 21 different emission lines. To make these calculations, the DAP first fits the stellar continuum using the Penalized Pixel-Fitting method \citep[pPXF, ][]{Cappellari_2004, Cappellari_2017} and then subtracts the best-fitting stellar continuum from the observed data before fitting single Gaussians to the emission lines, allowing for additional subtraction of a non-zero baseline. The final fitting model, emission line fit, and baseline fit are all available via the `MODEL,' `EMLINE,' and `EMLINE\_BASE' extensions respectively in the DAP logcube files. 

\subsection{Samples}
The work in this paper is based on the Data Release 14 (DR14) which consists of data cubes for 2778 galaxies for 2727 unique objects. The aim of this work is to compare the kinematic characteristics of the [OIII] emission line for MaNGA-selected AGN and non-AGN in the MaNGA sample. In optical surveys, emission line flux ratios and diagnostic diagrams are the most common way to identify AGN \citep{bald81,oste89,zaka03,kauf03a,kewl06,reye08,yuan16}. But a major caveat of large optical spectroscopic surveys such as the Sloan Digital Sky Survey (prior to MaNGA) is the small size of the optical fibres which, at 3$\arcsec$ diameter (in the case of SDSS-I to SDSS-III surveys), only cover a fraction of the footprint of a galaxy and are only sensitive to processes close to the galactic center.

\cite{wyle18} recently developed spatially resolved techniques for identifying signatures of active galactic nuclei tailored to MaNGA IFU data identifying 303 AGN candidates. A minor update to the selection code \footnote{This minor update regards the inclusion of a few `borderline' objects related to the precision with which $d_{BPT}$, the distance between Seyfert/LINER-classified spaxels in the BPT diagram and the star formation demarcation line, is measured (see \citet{wyle18} for more details).} has increased the sample to 308 sources which we adopt as the `AGN' sample in this work (see Table \ref{table_measurements}). We furthermore refer to remaining MaNGA galaxies as `non-AGN'. While LINER-like (`low ionisation nuclear emission line region') galaxies can be associated with a number of ionisation mechanisms such as weakly ionising AGN \citep{heck80}, shock ionisation (either related to star-forming processes in inactive galaxies or AGN activity) or photo-ionisation through hot evolved stars, the algorithm developed by \citet{wyle18} was tailored to select the most likely AGN among LINER-like galaxies. This has been achieved using combination of spatially resolved line diagnostic diagrams, assessing the significance of the deviation from the star formation locus in line diagnostic diagrams and applying additional cuts on H$\alpha$ surface brightness and H$\alpha$ equivalent width.

Some of the AGN candidates selected by \citet{wyle18} would not have been identified based on the single-fibre nuclear spectra alone. This is either because the AGN is hidden behind large columns of dust in the galactic center, because the AGN has recently turned off and relic AGN signatures are only visible at larger distances, because the AGN is offset after a recent galaxy merger, or because a circumnuclear starburst overwhelms nuclear AGN signatures \citep{wyle18}. Based on the MaNGA measurements in the inner 3$\arcsec$ (similar to the single fibre classifications of SDSS I-III), 109 out of the 308 MaNGA-selected AGN would be classified as star-forming (SF) galaxies, 84 sources would be classified as `Seyfert' galaxies, i.e. AGN, and 91 sources would be classified as LINER-like galaxies. In the remaining part of the paper, we refer to these subsamples as Seyfert-AGN (84 sources), SF-AGN (109 sources) and LINER-AGN (91 sources).
The 24 remaining galaxies could not be classified based on their central emission line signatures due to one or more needed emission lines not fulfilling the required S/N criteria \citep[see ][for the details of the selection]{wyle18}. In the remaining part of the paper, we do not include these galaxies when assessing the differences between the individual `types' of AGN.

We note that this AGN selection is different to the ones used in \citet{rembold17} or \citet{Sanchez18} who both present results on AGN in MaNGA. The main difference in the sample selection is that both of these works only use the photoionisation signatures in the central region of the galaxies to classify a galaxy as an AGN candidate. \citet{rembold17} uses the SDSS-III spectroscopic data from DR12 based on the 3\arcsec\ single-fibre measurements whereas \citet{Sanchez18} uses the MaNGA spectroscopic information in the central $3\arcsec \times 3\arcsec$ region analysed and measured with the PIPE3D pipeline \citep{Sanchez_2016}. Their AGN samples contain 62 and 98 AGN candidates, respectively. The differences in the sample selection lie in the exact choice of diagnostic diagrams and equivalent width cuts. For example, while \citet{rembold17} employ a cut of 3\AA\ on the equivalent width of H$\alpha$, \citet{Sanchez18} use a more relaxed criterion of only 1.5\AA. These differences lead to different and less/more sources being selected as AGN candidates. The overlap between the \citet{wyle18} and \citet{rembold17} samples is 37 sources, the overlap between the \citet{wyle18} and \citet{Sanchez18} samples is 44 sources and the overlap between the \citet{rembold17} and \citet{Sanchez18} samples is 37 sources. In Comerford et al. (submitted), we explore the overlap between optical and radio selected AGN in MaNGA and are furthermore exploring the implications for BPT and H$\alpha$ selection of AGNs in a subsequent paper (Negus et al., in prep.).

\section{Methods}

\subsection{Spectral Fitting}
In this paper, all kinematic calculations are made based on fits of the emission lines of the [OIII] doublet line at 5008/4960\angstrom. All SDSS data, including MaNGA, are stored at vacuum wavelengths \citep{Morton_1991} but we use air wavelengths to identify emission lines following the long-standing convention. In AGN, a large fraction of the [OIII] emission originates in the narrow-line regions surrounding the AGN \citep[e.g.][]{kewl06, liu13b}. Being a forbidden line, it can therefore trace the low-density AGN-ionised gas even out to galaxy-wide scales of several kiloparsecs \citep{liu13b}. [OIII] is easily observable from the ground for low to intermediate redshift AGN and it is widely used as a gas (outflow) tracer in low- and intermediate-redshift AGN \citep{cren10a, cren15, Lena_2015, Fischer_2017, rupk17}, making our measurements easily comparable with other works.

We develop a customised fitting procedure to model the [OIII] doublet and potential secondary and/or broad components in the line. We first extract the spectra for each spaxel from the DAP Logcube files using the `FLUX' extension, providing the flux density in units of  $  \frac{10^{-17} \rm{erg}}{\rm{s\ cm^2\ {\angstrom}\ \rm{spaxel}}}$ and then subtract the modelled stellar continuum (see Section 2.2) using the other Logcube extensions. We then also measure the flux-level blue- and red-ward of the H$\beta$ + [OIII] line complex to subtract any additional continuum contributions using a linear function of wavelength that might not have been accounted for. We further adopt the spectroscopic redshifts from the NASA Sloan Atlas (NSA) catalogues that are based on the single-fibre measurements to correct the spectra to the rest-frame of the galaxy. 

\begin{figure}
\centering
\includegraphics[width = 0.4\textwidth, trim = 4cm 3.2cm 6cm 2cm, clip= true]{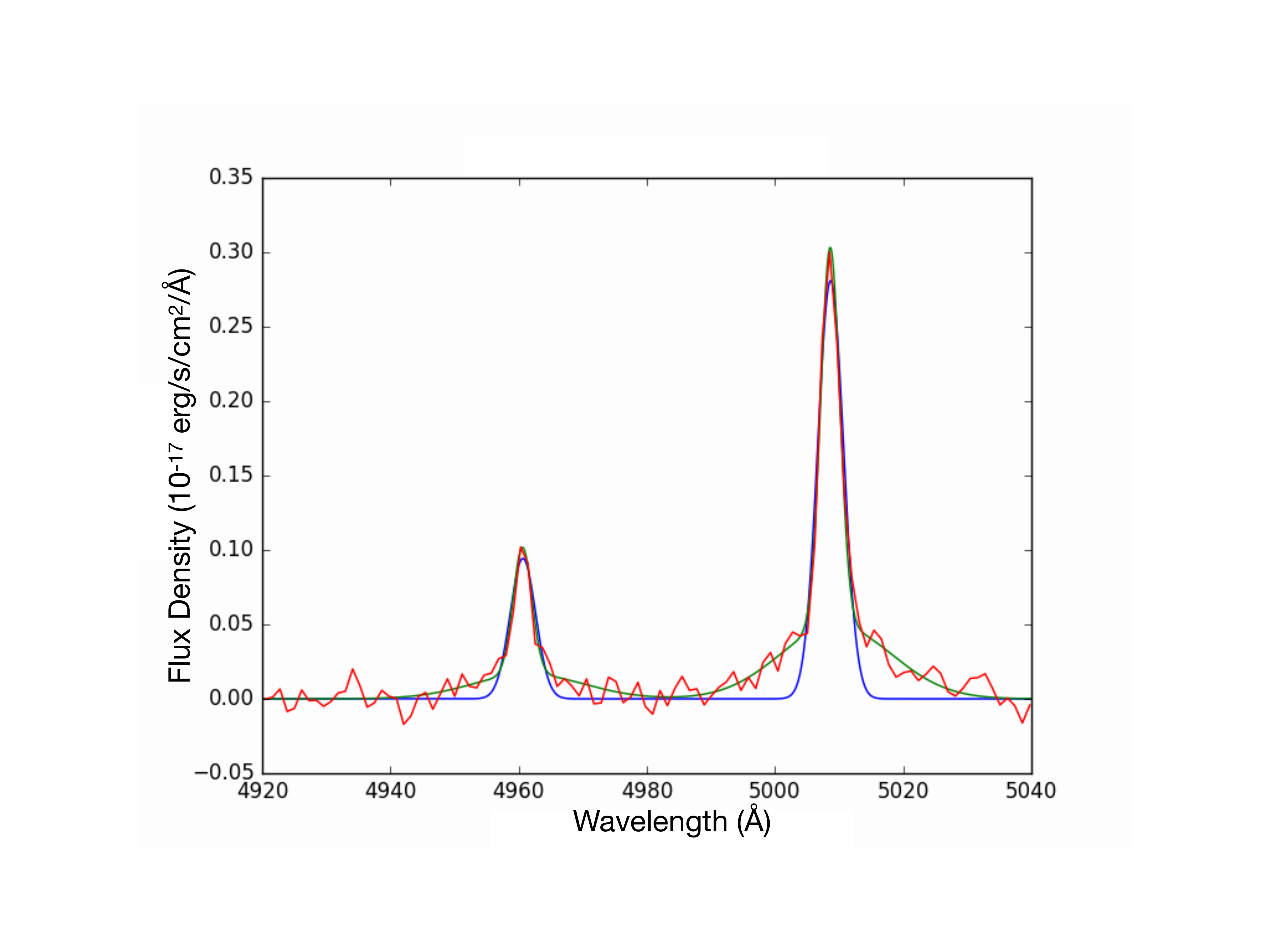}
\caption{An example fit for a spaxel in MaNGA galaxy 8715-3702. The spectrum after subtraction of the stellar continuum is shown in red, the single gaussian fit is shown in blue, and the multi-Gaussian fit is shown in green. While the single Gaussian fit (blue) misses the broad wings in the line profile, the multi-Gaussian fit provides a much more accurate description of the line profile and its associated kinematic properties.}
\label{8715-3702_fit}
\end{figure}

The MaNGA Data Analysis pipeline performs single Gaussian fitting on a number of selected, bright emission lines in all spaxels. However, a single Gaussian is often insufficient for describing the profile of the [OIII] line, and in particular such a fit would fail to capture a secondary broad component characteristic of outflowing gas. To evaluate the prevalence of additional kinematic components in MaNGA-selected AGN, we therefore allow multiple Gaussian components to be fit to the emission lines. In the fitting process, we fit the two transitions of the [OIII] doublet simultaneously and assume they share the same kinematics. We furthermore fix the ratio between the amplitude of the 5008\angstrom\ peak and the 4960\angstrom\ peak to its quantum value of 2.98. The fitting procedure uses least squares regression to return best-fit parameters for the single-Gaussian and double-Gaussian models. We evaluate the goodness of the fit based on its $\chi^2$ value and use the fit that both minimises the number of Gaussian components and its $\chi^2$. While in the 2-dimensional maps we show all spaxels where the signal-to-noise ratio of the [OIII] emission lines $S/N > 3$, we only use spaxels with $S/N > 10$ in the subsequent analysis part of this paper. In Figure \ref{8715-3702_fit} we show an example fit to a spaxel in the MaNGA galaxy 8715-3702 where our multiple component Gaussian fitting describes the line profile more accurately than the standard single Gaussian fit.

\subsection{Non-Parametric Values}


	When spectra can be well described based solely by single Gaussian fits, then the calculation of best fit parameters such as velocity dispersion $\sigma$, full width at half maximum (FWHM), amplitude are sufficient to describe the kinematic properties of the emission line in that spaxel. This is not the case when multiple Gaussians are used to describe the line profile. Because the sum of multiple gaussians is used in some spaxels, we calculate non-parametric values based on percentages of the total integrated flux and follow the measurement strategy presented in \citet{zaka14} and \citet{liu13b} \citep[see also][]{whit85a} to determine amplitudes, centroid velocities and emission line widths. Such non-parametric measurements do not strongly depend on a specific fitting procedure. 

The cumulative flux as a function of velocity is 

\begin{equation}
\Phi(v) = \int_{-\infty}^{v} F_{v}(v') dv'
\end{equation}
and the total line flux is given by $\Phi(\infty)$. In practice, we use the interval [-2000,2000]~km~s$^{-1}$ in the rest-frame of the galaxy for the integration. For each spaxel, we compute the line of sight velocity $v_{med}$ where $\Phi(v_{med}) = 0.5 \cdot \Phi(\infty)$, i.e. this is the velocity that bisects the total area underneath the emission-line profile. Because the fitting is performed in the rest frame of the galaxy as determined by its stellar component, $v_{med}$ is measured relative to the restframe. We use the W$_{80}$ parameter to parameterise the velocity width of the line. W$_{80}$ refers to the velocity width that encloses 80\% of the total flux. For a purely Gaussian profile, W$_{80}$ is close to the FWHM but the non-parametric velocity width measurements are more sensitive to the weak broad bases of non-Gaussian emission line profiles \citep{liu13b}. We first determine $v_{90}$ such that $\Phi(v_{90}) = 0.9 \cdot \Phi(\infty)$ and $v_{10}$ such that $\Phi(v_{10})~=~0.1~\cdot~\Phi(\infty)$ and then calculate W$_{80}$ using W$_{80} = v_{90} - v_{10}$.

\subsection{Kinematic and Division Maps}

Having performed the fitting and analysis procedure described in Section 3.1 and 3.2 in all spaxels of MaNGA-selected AGN candidates, we create two-dimensional maps for the following quantities: 
\begin{enumerate}
\item The total flux measured for [OIII]5008\angstrom, 
\item The non-parametric line-of-sight velocity $v_{med}$,
\item The non-parametric velocity width W$_{80}$,
\item The number of Gaussians used for each fit determined based on the $\chi^2$ analysis,
\item The reduced $\chi^2$ statistic of the best fit in each spaxel.
\end{enumerate}

\begin{figure*}
\begin{center}
\includegraphics[width = 0.75\textwidth, trim = 0cm 0.2cm 0cm 0cm, clip= true]{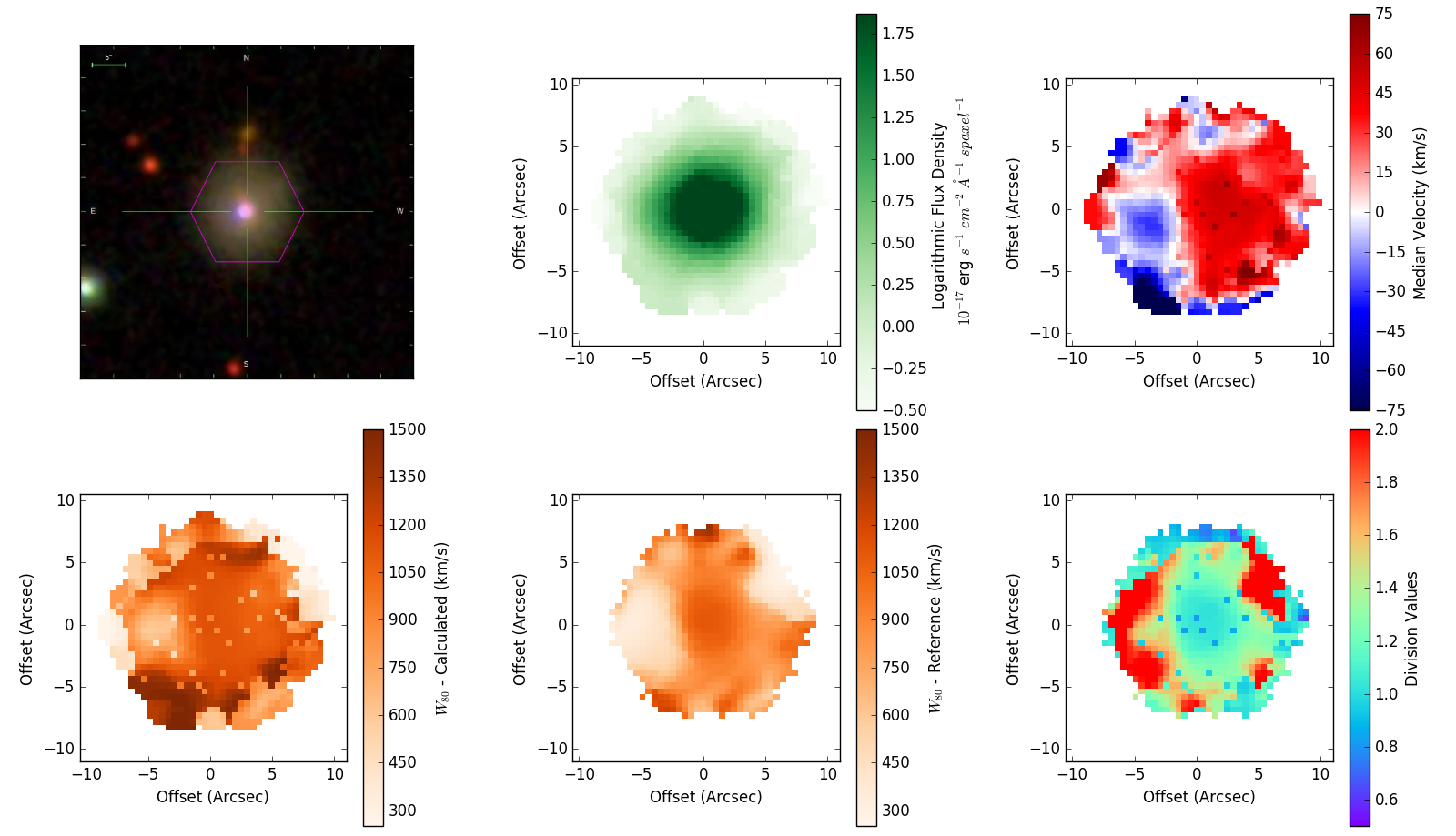}

\vspace{0.5cm}
\includegraphics[width = 0.75\textwidth, trim = 0cm 0.2cm 0cm 0cm, clip= true]{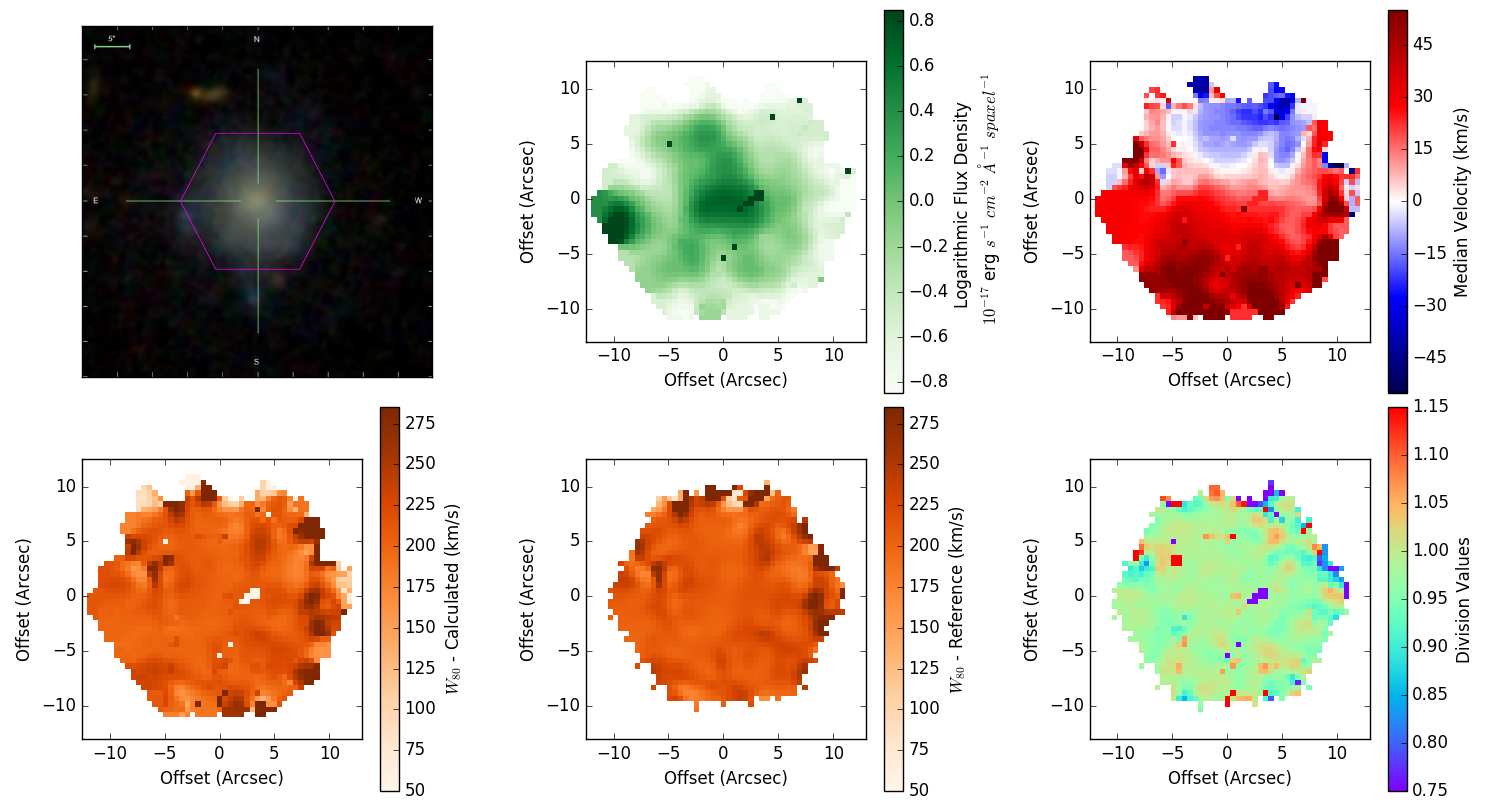}

\vspace{0.5cm}
\includegraphics[width = 0.75\textwidth, trim = 0cm 0.2CM 0cm 0cm, clip= true]{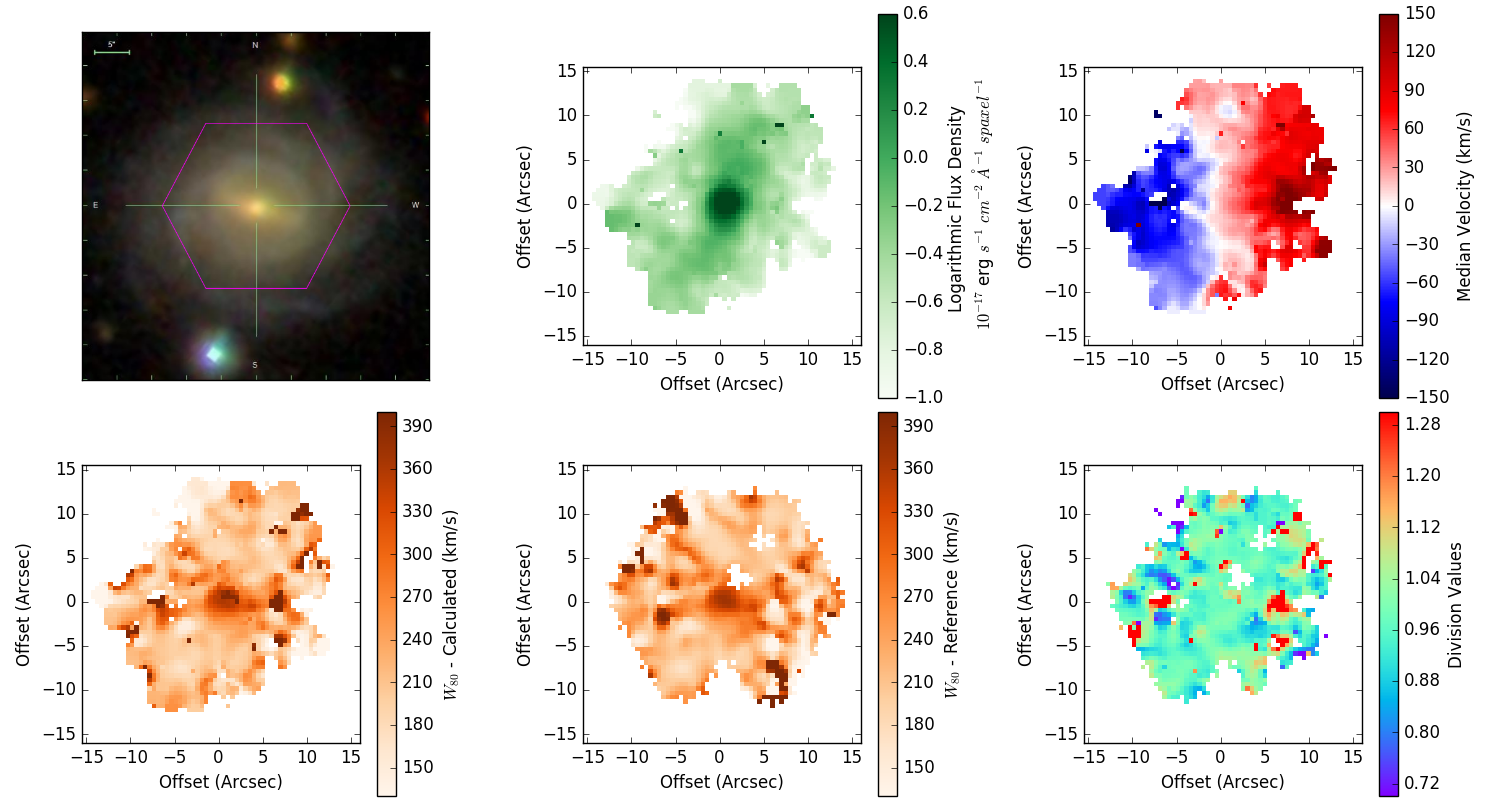}

\caption{Example maps for three MaNGA galaxies 8715-3702, 8459-6102 and 8978-9101. We show the SDSS composite $gri$ optical (top left), the [OIII] flux density (Logarithmic, top center), the median velocity $v_{med}$ (top right), the multi-Gaussian-based W$_{80}$ measurements (lower left), the reference, single-Gaussian-based W$_{80, DAP}$ values based on the MaNGA Data Analysis Pipeline fits (lower center), and the divisional values $\frac{\rm{W}_{80}}{\rm{W}_{80,DAP}}$(lower right). All images and maps are orientated North-up, East-left.}
\label{maps}
\end{center}
\end{figure*}

The MaNGA survey team has already made velocity dispersion calculations $\sigma_{DAP}$ based on a single gaussian fits to the [OIII] emission lines in each spaxel. In order to assess how well these single Gaussian fits describe the lines and how the velocity dispersion based on the single Gaussian fits compares to the dispersion derived in this work, we generate `division maps' for every galaxy in the survey. The `division maps' report the fractional difference in velocity dispersion when comparing the single Gaussian fits to our measurements. To achieve a fair comparison, we first compute the non-parametric velocity dispersion W$_{80}$ for the single Gaussian fits from the MaNGA DAP. As mentioned above, for a purely Gaussian profile, W$_{80}$ is closely related to the FWHM and therefore to its velocity dispersion, such that:

\begin{equation}
\centering
W_{80, \rm{DAP}} = 1.088 \cdot FWHM_{\rm{DAP}} = 1.088 \cdot 2.35 \sigma_{\rm{DAP}} = 2.56 \cdot\sigma_{\rm{DAP}} 
\label{convert}
\end{equation}

We then divide the W$_{80}$ value derived from our multi-Gaussian customised fitting procedure described above by the corresponding DAP single-Gaussian based W$_{80, \rm{DAP}}$ value in each spaxel. This results in a two-dimensional map for each galaxy reporting divisional values (i.e. fractional differences in velocity line width measurements) across the galaxy. Values in this map $\sim 1$ identify the spaxels where our fits, either single or double, are similar to the DAP ones, while higher values flag the spaxels where the double Gaussian fit captured a secondary component in the emission line missed by the DAP. In the subsequent analysis, we utilise these maps to assess the prevalence of additional kinematic ionised gas components in MaNGA galaxies and MaNGA-selected AGN. 

In Figure \ref{maps} we show three examples of these maps, including the galaxy 8715-3702 already shown in Figure \ref{8715-3702_fit}. For each source, we show the SDSS composite image, the
[OIII] flux density, the median velocity $v_{med}$, the multi-Gaussian-based W$_{80}$ measurements, the single-Gaussian-based W$_{80,\rm{DAP}}$ and the divisional values. We note that the number of spaxels with valid values might differ between the W$_{80,\rm{DAP}}$ map and the maps based on the here developed fitting routine. This is because we are only reporting measured quantities in pixels with a S/N of the [OIII] emission line of $> 3$ which might differ slighly from the DAP `good' spaxels. Object 8715-3702 (top source) is classified as a Seyfert-AGN with high W$_{80}$ measurements of $\sim 1000$~km/s while the W$_{80, \rm{DAP}}$ measurements are significantly lower. This difference is reflected in the `division map' where the enhanced velocity line width in the North-West and East of the galaxy is apparent. Objects 8459-6102 (centre source, classified as a regular star-forming galaxy) and 8978-9101 (bottom source, classified as LINER-AGN) both show a low gas velocity width and little difference between the W$_{80}$ and W$_{80, \rm{DAP}}$ measurements. That means that most fits are consistent with the single-Gaussian fit results suggesting that no or little enhanced gas kinematics are present in these sources.

\section{Results}

\subsection{Absolute Kinematic Comparison}

\begin{figure}
\includegraphics[ width=0.45\textwidth, trim = 0cm 0.2cm 1.3cm 1.54cm, clip= true]{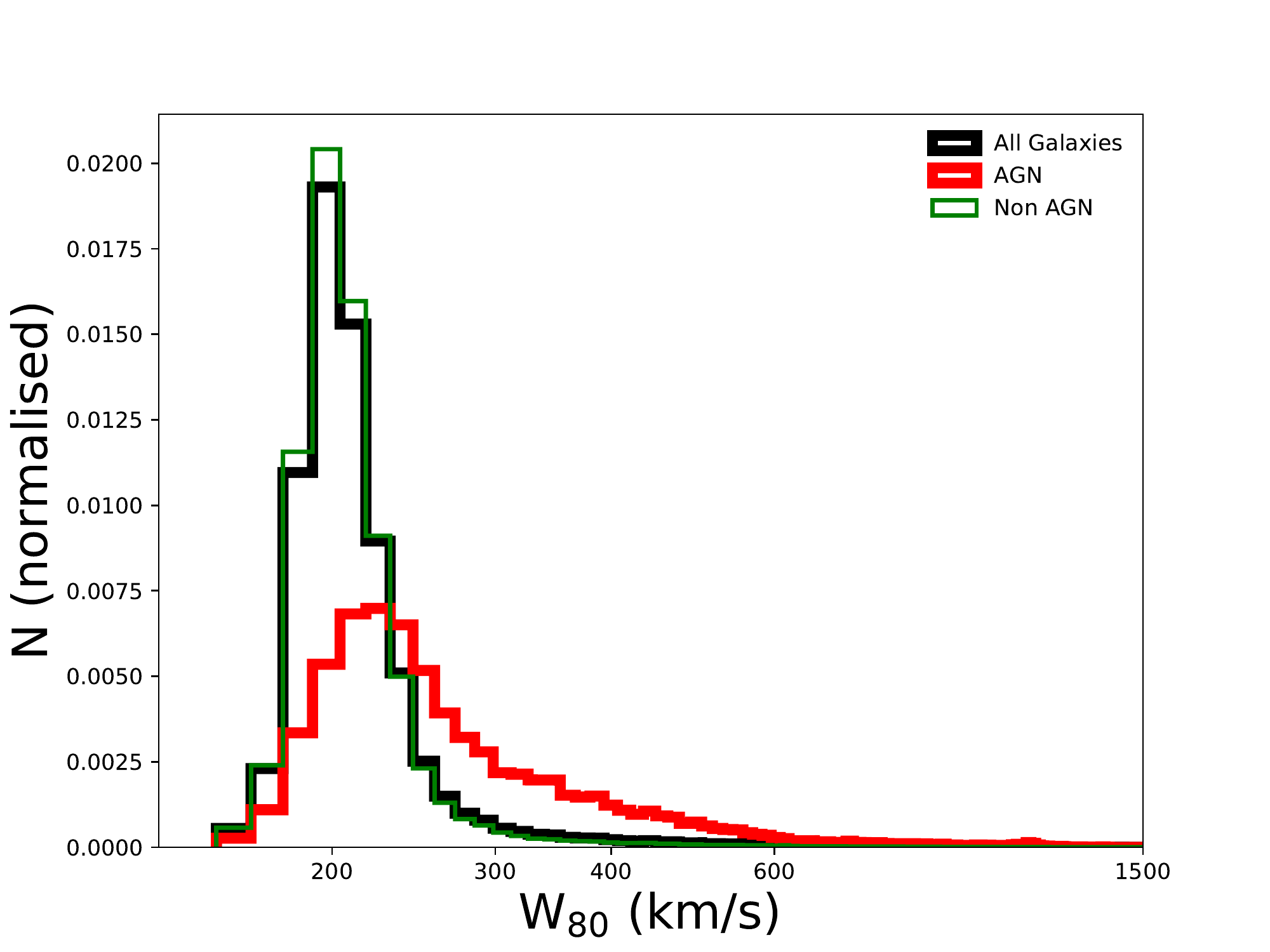}
\caption{The distribution of W$_{80}$ measurements across all spaxels in all 2778 MaNGA galaxies (black), the MaNGA-selected AGN (red) and non-AGN (green). The small number of values below 150 km/s are removed as they are at the limit of the instrumental resolution of the survey and not physically meaningful. Of particular interest is the largely skewed tail to values above \textasciitilde500 km/s, which is significantly enhanced for the MaNGA-selected AGN ($p-value$ of KS-test $< 10^{-200}$), indicative of enhanced kinematics potentially related to current or previous AGN activity.}
\label{big_hist}
\end{figure}

We first perform an absolute kinematic comparison by assessing the distribution of W$_{80}$ values in all MaNGA galaxies and in the AGN samples. Figure \ref{big_hist} shows the distribution of all W$_{80}$ measurements in every fitted spaxel of every MaNGA galaxy. We also separately show the distributions of the MaNGA-selected AGN and in non-AGN galaxies. The distribution peaks at $\sim 200$~km~s$^{-1}$ which corresponds to typical gas velocity widths in galaxies with masses of $10^{10-11}$~M$_{\odot}$. Additionally, we observe a heavily skewed tail towards large velocity widths, which is enhanced in MaNGA-selected AGN, indicative of kinematic peculiarities \citep{nels00, kewl06}.

\begin{figure*}
\begin{center}
\includegraphics[width=0.95\textwidth, trim = 4cm 1cm 4cm 2cm, clip= true]{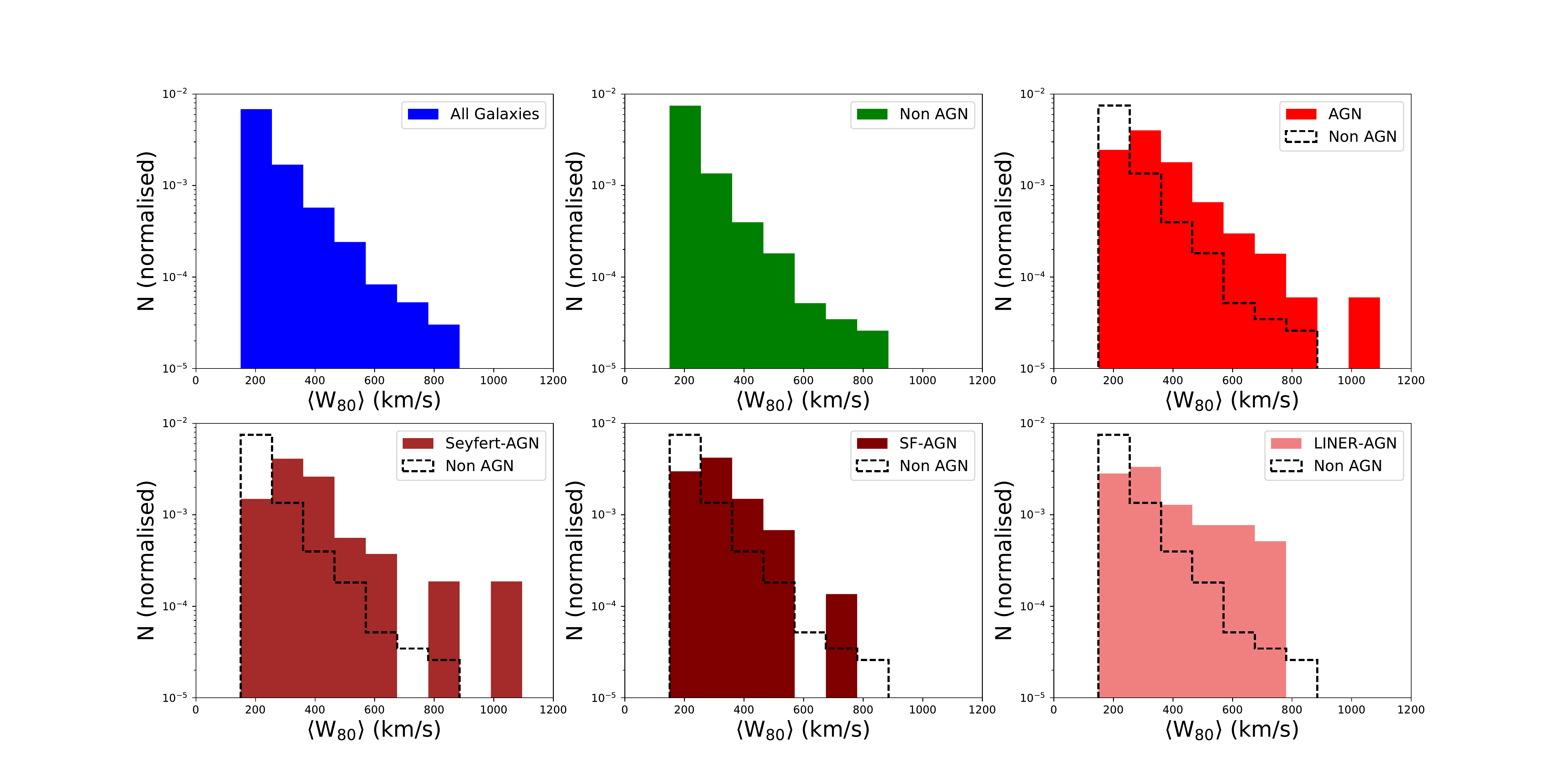}
\caption{Normalised logarithmic distributions of the mean W$_{80}$ measurements for each galaxy for the whole MaNGA sample, the MaNGA-selected AGN, the AGN subsamples and the non-AGN. The small number of values below 150 km/s are removed as they are at the limit of the instrumental resolution of the survey and not physically meaningful. We show the distribution for all MaNGA galaxies (top left), non-AGN (top center), AGN Candidates (top right), and the AGN subsamples Seyfert-AGN (lower left), SF-AGN (lower center), and LINER-AGN (lower right). For reference, we also show the non-AGN distribution (black dashed histogram) in the AGN and AGN subsample panels. In the AGN and all AGN subsamples, the distributions show larger contributions from high W$_{80}$ measurements, indicative of enhanced kinematics potentially related to current or previous AGN activity.}
\label{W80_hist_mean}
\end{center}
\end{figure*}

In Figure \ref{W80_hist_mean} we show the distribution of the mean W$_{80}$ measurements $\langle$W$_{80}\rangle$ for all galaxies, for the MaNGA-selected AGN and the three different MaNGA-AGN subsamples, SF-AGN, LINER-AGN and Seyfert-AGN. For this analysis we only include galaxies in which at least 10\% of the spaxels have valid [OIII] emission line measurements, i.e. a peak $S/N > 10$. This cut was chosen to ensure that [OIII] is  detected in a significant enough number of spaxels and to be able make meaningful conclusions about the [OIII] behaviour in these galaxies. In total, 1116 MaNGA galaxies and 159 MaNGA-selected AGN fulfill this criterion.

\begin{figure*}
\begin{center}
\includegraphics[width=0.95\textwidth, trim = 4cm 1cm 4cm 2cm, clip= true]{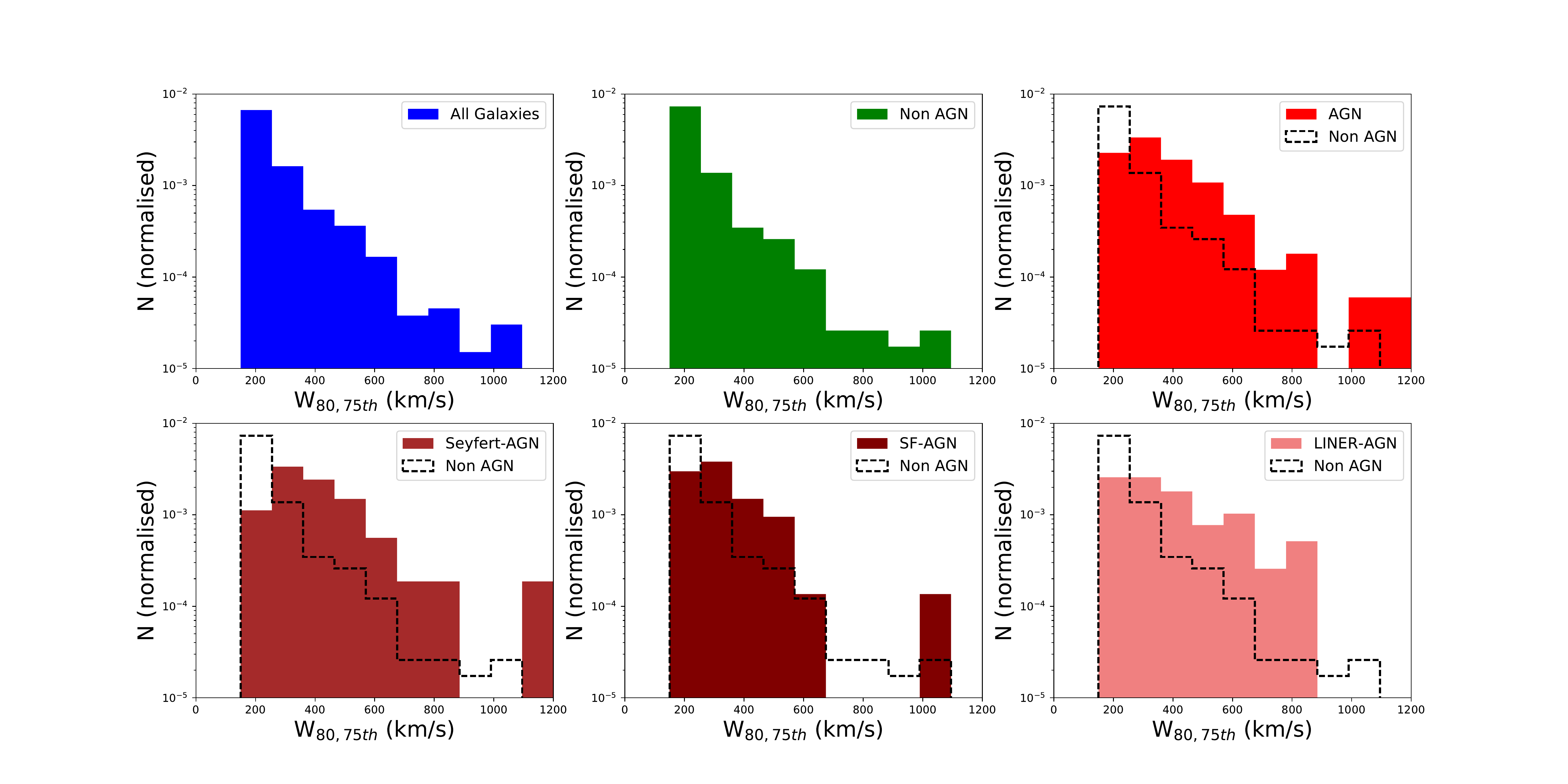}
\caption{Normalised logarithmic distributions of the 75th percentile W$_{80}$ measurements for each galaxy for the whole MaNGA sample, the MaNGA-selected AGN, the AGN subsamples and the non-AGN. The small number of values below 150 km/s are removed as they are at the limit of the instrumental resolution of the survey and not physically meaningful. We show the distribution for all MaNGA galaxies (top left), non-AGN (top center), AGN Candidates (top right), and the AGN subsamples Seyfert-AGN (lower left), SF-AGN (lower center), and LINER-AGN (lower right). For reference, we also show the non-AGN distribution (black dashed histogram) in the AGN and AGN subsample panels. In the AGN and all AGN subsamples, the distributions show larger contributions from high W$_{80}$ measurements, indicative of enhanced kinematics potentially related to current or previous AGN activity.}
\label{W80_hist_75}
\end{center}
\end{figure*}

For every galaxy, we furthermore measure the 75th percentile of their W$_{80}$ distributions W$_{80, 75th}$. In Figure \ref{W80_hist_75} we show the distribution of the W$_{80, 75th}$ measurements for all galaxies, for the MaNGA-selected AGN and the three different MaNGA-AGN subsamples, SF-AGN, LINER-AGN and Seyfert-AGN. 

A two-sample Kolmogorov-Smirnov (KS) test shows that the $\langle$W$_{80}\rangle$ distributions of the MaNGA-selected AGN, the SF-AGN, LINER-AGN and Seyfert-AGN are significantly different from the total distribution. We repeat the analysis comparing the distribution of the MaNGA-selected AGN subsamples with only the non-AGN MaNGA galaxies and find similarly significant results. We report the measurements for $\langle$W$_{80}\rangle$ and W$_{80, 75th}$ in Table \ref{table_measurements} and the $p$-statistic value of the KS-tests in Table \ref{table_statistics}. To investigate if these results are driven by any unaccounted bias, we randomly select 100 galaxies from the full sample and repeat the statistical comparison. A KS-test comparing the full and random galaxy sample results in a high $p$-value of $p = 0.47$ and shows that the W$_{80}$ distribution of the randomly selected galaxy sample is drawn from the same distribution as the full MaNGA sample. We conclude that the statistical difference between the AGN and AGN-subsamples to the full MaNGA W$_{80}$ distribution is indeed intrinsic and that the kinematic properties of the MaNGA-selected AGN is distinct from the overall MaNGA distribution.

To further illustrate this point, we compute the number of galaxies in which at least 10\% of the spaxels show W$_{80}$ values of W$_{80} > 500/800/1000$~km~s$^{-1}$. A total of 257/112/37 galaxies pass this cut. In the MaNGA-selected AGN sample, 77/21/7 (25/7/2\%) of the galaxies pass this cut while only 180/91/30 (7/4/1\%) of the remaining MaNGA galaxies do. Similarly, there are 13 MaNGA-selected AGN with $\langle$W$_{80}\rangle > 500$~km~s$^{-1}$ (4\%) and only 20 non-AGN with $\langle$W$_{80}\rangle > 500$~km~s$^{-1}$ (< 1\%). The fractions were computed using the full MaNGA and MaNGA AGN sample as baseline, i.e. 2778 and 308 sources, respectively. Using the number of galaxies that initially passed our quality cut (at least 10\% of the spaxels need to have an [OIII] line measurement with S/N>10) as baseline, i.e. 1116 and 159 sources, respectively, leads to the same conclusion: Two to three times as many MaNGA selected AGN show enhanced [OIII] kinematics compared to the non-AGN in MaNGA.

{\footnotesize
\begin{table*}
\caption{Mean and 75th percentile [OIII] velocity width measurements $\langle$W$_{80}\rangle$ and W$_{80, 75th}$ of all MaNGA galaxies analysed in this work. The column  f$_{spx}$ reports the fraction of spaxels with high signal-to-noise S/N [OIII] emission line measurements with S/N~$> 10$. The last two columns report whether the source is identified as a MaNGA-selected AGN (flag1 = 1), a MaNGA non-AGN (flag1=0). If identified as a MaNGA-selected AGN, the flag2 column denotes whether this source is a SF-AGN (flag2=2), LINER-AGN  (flag2=3) or Seyfert-AGN (flag2=4, see Section 2.2 for more details on the subsample definitions). }
\begin{tabular}{|l|r|r|r|r|r|r|r|r|}
\hline
  \multicolumn{1}{|c|}{MaNGA ID} &
  \multicolumn{1}{c|}{R.A.} &
  \multicolumn{1}{c|}{Dec.} &
  \multicolumn{1}{c|}{z} &
  \multicolumn{1}{c|}{$\langle$W$_{80}\rangle$} &
  \multicolumn{1}{c|}{W$_{80, 75th}$} &
  \multicolumn{1}{c|}{f$_{spx}$} &
  \multicolumn{1}{c|}{flag1} &
  \multicolumn{1}{c|}{flag2} \\
   \multicolumn{1}{|c|}{} &
  \multicolumn{1}{c|}{(degrees)} &
  \multicolumn{1}{c|}{(degrees)} &
  \multicolumn{1}{c|}{} &
  \multicolumn{1}{c|}{(km/s)} &
  \multicolumn{1}{c|}{(km/s)} &
  \multicolumn{1}{c|}{} &
  \multicolumn{1}{c|}{} &
  \multicolumn{1}{c|}{} \\
\hline
  1-593159 & 217.629970676 & 52.7071590288 & 0.0825072 & 375 & 419 & 0.26 & 1 & 2\\
  1-592984 & 215.718400019 & 40.6225967971 & 0.0922799 & 198 & 207 & 0.15 & 0 & 0\\
  1-592881 & 214.647824001 & 44.1224317199 & 0.0220107 &  &  & 0.0 & 0 & 0\\
  1-592881 & 214.647824001 & 44.1224317199 & 0.0362875 &  &  & 0.0 & 0 & 0\\
  1-592743 & 213.110482013 & 45.6904101161 & 0.0311029 & 224 & 228 & 0.87 & 0 & 0\\
  1-592049 & 206.299565786 & 23.0718639552 & 0.0189259 &  &  & 0.0 & 0 & 0\\
  1-591474 & 197.580687349 & 47.1240556946 & 0.0441288 & 295 & 325 & 0.04 & 1 & 3\\
  1-591248 & 195.01782 & 28.15501 & 0.0305763 &  &  & 0.0 & 0 & 0\\
  1-591183 & 194.806227507 & 27.7746020056 & 0.0416588 &  &  & 0.0 & 0 & 0\\
  1-591068 & 194.181347734 & 27.0347535248 & 0.0314488 & 250 & 254 & 0.14 & 0 & 0\\
  1-591006 & 193.579232526 & 27.0680204421 & 0.13901 &  &  & 0.0 & 0 & 0\\
  1-590159 & 187.063503724 & 44.4531321941 & 0.0489803 & 464 & 560 & 0.20 & 1 & 2\\
  1-590053 & 186.11866 & 46.01868 & 0.0435213 & 210 & 210 & 0.68 & 0 & 0\\
  1-589908 & 184.55356586 & 44.1732422277 & 0.127436 & 331 & 332 & 0.41& 0 & 0\\
  \hline\end{tabular}
  \\
  Only a portion of this table is shown here to demonstrate its form and content. A machine-readable version of the full table is available as online material.
  \label{table_measurements}
  \end{table*}
}

\subsection{Proportional Comparison}

While the `Absolute Kinematic Comparison' presented above is sensitive to sources in which large values of W$_{80}$ are observed, many outflow candidates and kinematically peculiar sources would not be selected. In low- and intermediate luminosity AGN driving outflows, the measured gas velocity widths are often on the order of the velocity widths expected of regular disk rotation \citep{gree05o3, barb09, Fischer_2017, wyle17a}. In such sources, outflowing components can be better identified when broad, blue-shifted components are present in the relevant emission lines, [OIII] in our case. To quantify the prevalence of such components, we analyse the difference between the velocity width maps provided by the MaNGA Data Analysis Pipeline (based on single Gaussian fits) and the velocity width maps derived in this work using the `Division Maps' that report the fractional change between the pipeline velocity width and the here derived velocity width. 

\begin{figure*}
\begin{center}

\includegraphics[width = 0.95\textwidth, trim = 4cm 0cm 4cm 1cm, clip= true]{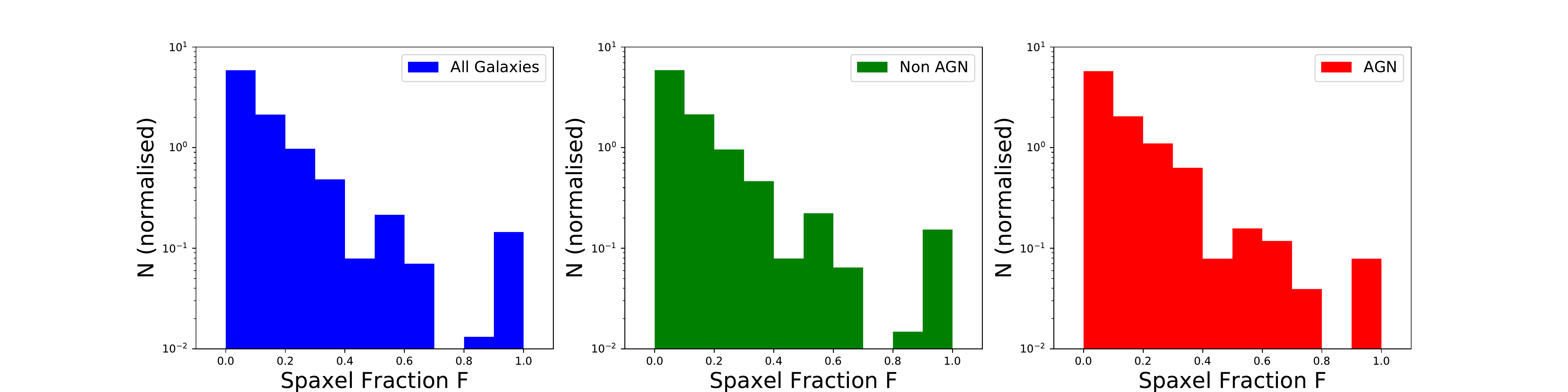}
\caption{Distribution of spaxel fractions for which the ratio between the single-Gaussian-based and multi-Gaussian-based velocity width exceeds a defined threshold $c$. We show the normalised distributions for all MaNGA galaxies (left), non-AGN (center) and MaNGA-selected AGN candidates (right) for a $c$ value of 1.25.}
\label{c plots}
\end{center}
\end{figure*}

In order to quantitatively assess the differences between the single-Gaussian-based and multi-Gaussian-based velocity widths, we measure the fraction of spaxels in each galaxy for which the ratio between the single-Gaussian-based and multi-Gaussian-based velocity width exceeds a defined threshold. We define $c$ as the threshold constant and show the distribution of spaxel fractions for $c = 1.25$ in Figure \ref{c plots}. The figure shows the distribution of the fraction of spaxels $F$ per galaxy with $\rm{W}_{80}> 1.25 \cdot \rm{W}_{80, DAP}$. Due to low number statistics we here do not repeat the analysis for the different AGN subsamples. The results of the two-sided KS-test comparing the overall, non-AGN and AGN distributions to that of the full MaNGA sample are reported in Table \ref{table_statistics}. The KS-test shows very low $p$-values when comparing the AGN distribution to the full MaNGA and/or non-AGN distribution, showing that the kinematic properties in these sources are distinct. Visually inspecting the distributions, we note that indeed most of the differences between the AGN distribution and the overall MaNGA distribution lies in the high spaxel fraction tail. This shows that not only does the distribution of the AGN sample differ from the overall MaNGA distribution, but that MaNGA-selected AGN show on average more spaxels where multiple Gaussian components were needed to describe the [OIII] emission line profile.

This trend is further quantitatively validated by introducing the following cut. We select galaxies where at least 25\% of the host galaxy's spaxels demonstrated at least a 25\% increase (this corresponds to $c = 1.25$) of their measured W$_{80}$ values compared to W$_{80, DAP}$. A 25\% increase is chosen to reflect areas of significant change, and 25\% of spaxels are required to limit the number of galaxies to a manageable amount while still identifying interesting sources. A total of 237 galaxies pass this cut out of 2778 total. 36 (11.6\%) of the galaxies in the AGN sample pass this cut compared to only 237 (9.5\%) of the remaining non-AGN.

To test for any potential biases in our analysis, we again draw a sample of 100 randomly selected galaxies from the overall MaNGA galaxy sample and repeat the analysis. A KS-test comparing the random distribution with the overall MaNGA distribution shows that the two distributions are statistically indistinguishable.

\begin{table*}
\caption{Results of the two-sided Kolmogorov-Smirnov test comparing the distributions of the `absolute kinematic comparison' shown in Figure~\ref{W80_hist_mean} and Figure~\ref{W80_hist_75} and the `proportional kinematic comparison' shown in Figure \ref{c plots}. We report the returned $p$-values when comparing the full MaNGA distribution and the non-AGN distribution to the AGN and AGN subsample distributions. Low $p$-values $< 0.01$ show that the distributions are significantly different from one another.}
\begin{tabular}{ l|c|c|c|c|c|c }
\hline
Absolute Kinematic Comparison (mean W$_{80})$: & & & & & \\
  & AGN & Seyfert-AGN & SF-AGN & LINER-AGN & Random \\
\hline
Full MaNGA & 1.0e-23 & 3.4e-12 & 6.8e-10 & 4.7e-7 & 0.47\\
non-AGN & 1.2e-31 & 2.4e-15 & 4.0e-13 & 1.5e-9 & 0.69\\
\hline 
 & & & & & \\
Absolute Kinematic Comparison (75th percentile W$_{80}$):\\
  & AGN & Seyfert-AGN & SF-AGN & LINER-AGN & Random \\
\hline
Full MaNGA & 4.9e-23 & 1.3e-11 & 4.2e-12 & 1.7e-7 & 0.71\\
non-AGN & 9.8e-31 & 2.4e-14 & 9.5e-16 & 5.9e-10 & 0.98\\

\hline 
& & & & & \\
Proportional Kinematic Comparison: & & & & & \\
  & AGN &  & &  & Random \\
\hline
Full MaNGA & 1.5e-4 &  &  &  & 0.30\\
non-AGN & 5.9e-6 &  &  &  & 0.52\\
\hline
\end{tabular}
\label{table_statistics}
\end{table*}

\section{Discussion}

\subsection{The prevalence of ionised outflow signatures in MaNGA galaxies}

With the goal of assessing the prevalence of ionised outflow signatures in MaNGA galaxies and MaNGA-selected AGN, we have shown that MaNGA-selected AGN candidates more frequently show enhanced [OIII] emission line kinematics than non-AGN in MaNGA. The difference in the gas kinematics between AGN and non-AGN galaxies is apparent in a variety of tests. 

By measuring the velocity width of the [OIII] emission line at $5007$\AA\ W$_{80}$, we have first shown that the $\langle$W$_{80}\rangle$ and W$_{80, 75th}$ distributions of the full MaNGA sample are significantly different from the MaNGA-selected AGN sample and from the individual AGN subsamples. MaNGA-selected AGN candidates tend to have higher $\langle$W$_{80}\rangle$ and W$_{80, 75th}$ and two to three times as many MaNGA-selected AGN candidates show enhanced [OIII] kinematics. In particular, we observe $2-3$ times as many of AGN with a significant fraction of spaxels with $\langle$W$_{80}\rangle > 500$~km/s compared to the non-AGN sample. These gas velocity values are significantly higher than what is expected from regular disk rotation. The typical gas velocity dispersion of SDSS galaxies at $z < 0.15$ and log$(M_{*}/M_{\odot}) < 11$ is $\lesssim 150$~km/s corresponding to W$_{80}$ values of $\lesssim 380$~km/s \citep{Thomas_2013, Beifiori_2014, Cicone_2016, Ubler_2019}. This suggests that a large fraction of the high velocity gas detected in the MaNGA-selected AGN is due to non-gravitational motions of the gas, potentially due to radiatively or mechanically AGN-driven outflows.

Most MaNGA-selected AGN are low-/intermediate-luminosity sources with $L_{[OIII]} \sim 10^{40}$erg/s \citep{wyle18}. Since the [O III] luminosity can be used as an indicator of AGN bolometric luminosity if an AGN is present in the galaxy \citep{heck04,reye08}, this corresponds to a bolometric luminosiy $L_{bol, AGN} \sim 10^{43}$erg/s. Recently, \citet{fior17} has collected AGN wind observations for nearly 100 AGN and shown that the outflow velocity strongly depends on the AGN bolometric luminosity confirming previous studies. In many intermediate luminosity AGN, such as the AGN in this work, the velocities of any AGN-driven outflow therefore tend to be low \citep[$v < 500$~km/s, ][]{Lena_2015, Fischer_2017, wyle17a, wyle18b} and single-Gaussian measurements and measuring the velocity width alone usually does not capture such low-velocity outflow activity. But additional kinematic components (such as outflows) often leave an imprint on the overall velocity profile of the relevant emission line ([OIII] in our case) which can be identified by multi-component analysis of the emission lines. By identifying MaNGA galaxies in which the [OIII] emission line is better described by a two-component fit -- compared to the single component fit performed by the MaNGA Data Analysis Pipeline -- we have shown that MaNGA-selected AGN require more often a multi-component model to describe their [OIII] emission line profiles suggesting an enhancement of broad secondary components being present.

The number of sources with outflow signatures identified in this paper is likely a lower limit. Since the analysis in this paper only captures sources with high [OIII] velocity widths or [OIII] velocity profiles with clear deviations from a Gaussian profile, weak and physically small outflow signatures below the spatial resolution of the MaNGA survey, such as the ones found in \cite{wyle17a}, would not be identified. A thorough modelling and subtraction of the gas velocity fields, which is beyond the scope of this paper, would be necessary to identify outflows through their residual signatures.

Our analysis shows that enhanced gas kinematics are clearly more prevalent in MaNGA-selected AGN than in non-AGN, but what are possible driving mechanisms? In low- and intermediate luminosity AGN, in which AGN-driven outflows are not expected to be of such high velocities that the AGN can be considered as the only possible driver, the effect of stellar feedback and merger-induced gas flows have to be considered as potentially contributing to the observed signatures. 

To examine the contribution of star formation to the outflow activity, we show the relation between $\langle$W$_{80}\rangle$ and the star formation rate (SFR) in Figure \ref{sfr_w80}. \citet{heck15} have shown that there is a strong correlation between the star formation rates and the outflow velocities of starburst-driven winds which can be explained by a model of a population of clouds accelerated by the combined forces of gravity and the momentum flux from the starburst. While \citet{heck15} measure the outflow velocity as the flux-weighted line centroid defined relative to the systemic velocity of the galaxy, a different definition (such as a W80 measurement) would not affect the qualitative sense of their results. 

We cross-match the MaNGA catalog with the MPA-JHU catalog and use the star formation rates reported there. The SFRs are derived using the 4000 \AA\ break of SDSS single fibre spectra, following the method described in \citet{Brinchmann_2004}. We do not observe any significant positive correlations between SFR and $\langle$W$_{80}\rangle$ for either the total MaNGA distribution or only the AGN subsample. Especially for the sources with $\langle$W$_{80}\rangle > 500$~km/s the star formation rates are not high enough to explain the observed high velocity widths. We also test whether there is any correlation between the specific star formation rates and $\langle$W$_{80}\rangle$ and do not observe any. 

We furthermore visually inspect all MaNGA-selected AGN with $\langle$W$_{80}\rangle > 500$~km/s and find the merger fraction to be $\sim 10$\%. We find a similar fraction for the total MaNGA AGN candidate sample. In this analysis we have included all galaxies in close pairs, interacting galaxies and galaxies with visible merger signatures such as tidal tails. \citet{rembold17} find an even lower merger fraction in their AGN host galaxy sample of only a few percent. 

While we cannot fully exclude that starbursts and mergers contribute partially to the here observed enhanced [OIII] kinematics, these results suggest that stellar-driven winds and merger-induced flows are not the dominant reason for why we observe high [OIII] velocity widths in the MaNGA-selected AGN candidates with $\langle$W$_{80}\rangle > 500$~km/s. Furthermore, spatially resolved inflows in isolated galaxies are usually associated with low velocity dispersions of a few tens to 100~km/s \citep{Storchi_2019}, such that it seems unlikely that the high velocity dispersions observed here would be related to any inflows. The signatures are consistent with being due to radiatively or mechanically-driven AGN outflows. This type of analysis is more difficult for the AGN outflow candidates that show [OIII] velocity profiles with clear deviations from a Gaussian profile (the ones identified in Section 4.2), but which do not necessarily have $\langle$W$_{80}\rangle > 500$~km/s. In the following section, we assess whether the kinematics of the additional kinematic components with which these profiles have been modelled agree with expectations from outflow models.

\subsection{Inflow vs. Outflow?}

Both types of comparisons presented in Section 4.1 and 4.2 reveal that the [OIII] kinematics of MaNGA-selected AGN are different from the non-AGN in MaNGA. In the previous section we have shown that the kinematics of the MaNGA AGN with $\langle$W$_{80}\rangle > 500$~km/s are consistent with being due to radiatively or mechanically-driven AGN outflows. In AGN with an enhanced prevalence of secondary kinematic components but not necessarily high [OIII] velocity widths, the kinematic signatures could be due to either outflows or inflows. In this section we determine which scenario is more plausible.

\begin{figure*}
\begin{center}
\includegraphics[scale = 0.45, trim = 0cm 0cm 0cm 0cm, clip= true]{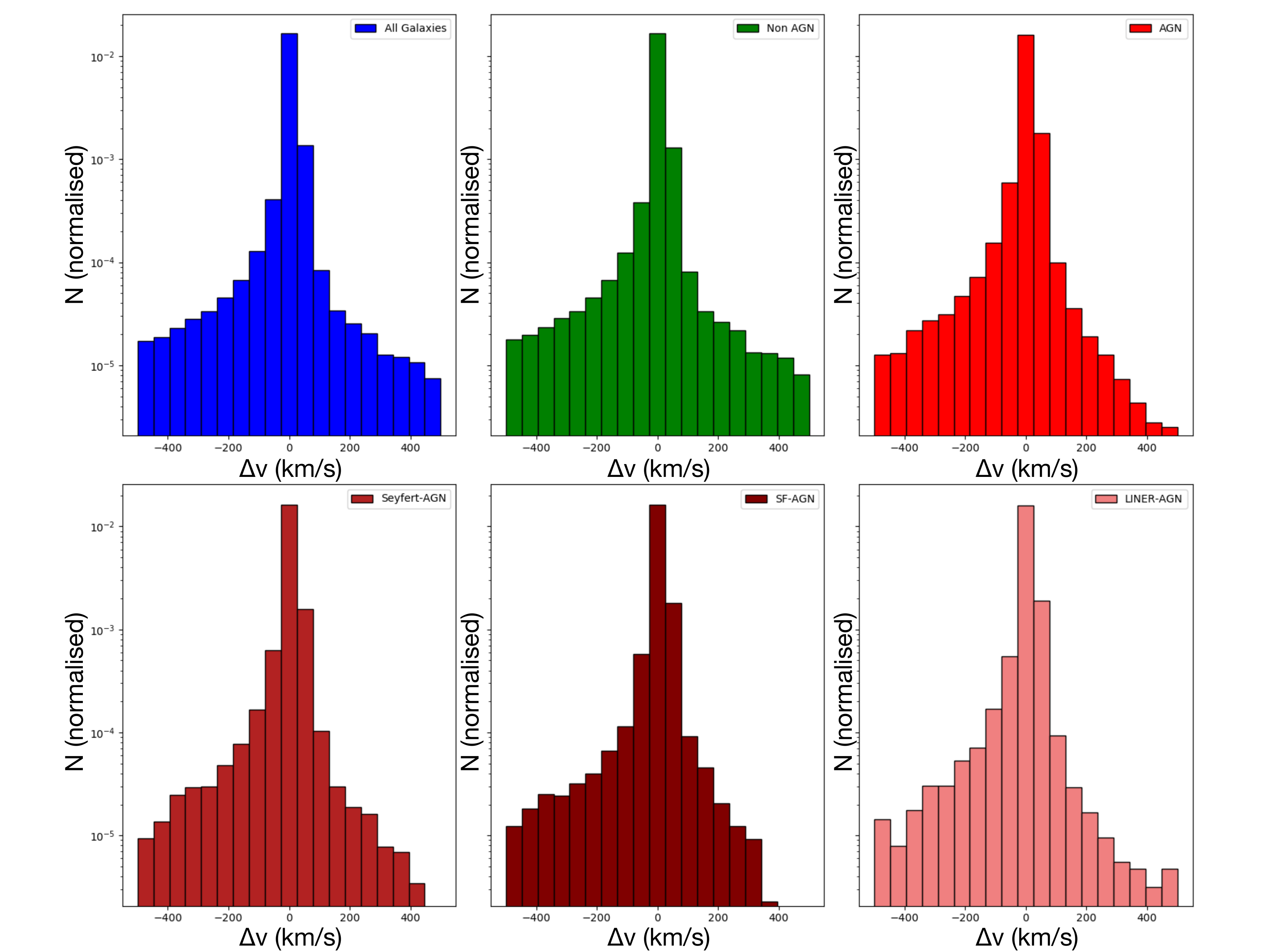}
\caption{Distribution of the [OIII] velocity offsets $\Delta v$, measuring the difference between the Gaussian-measured line-of-sight velocity and the non-parametric line-of-sight velocity as determined from our multi-Gaussian fitting and indicating whether the emission line is blue- or redshifted with respect to the Gaussian-measured line-of-sight velocity.  We show the distributions for all galaxies in MaNGA (upper left), the non-AGN in MaNGA (upper center), the MaNGA-selected AGN candidates (upper right) and the three AGN subsamples (lowwo ist er row). The distribution of the MaNGA-selected AGN candidates and the three AGN subsamples are skewed towards negative $\Delta v$ values, suggesting that we predominantly observe the signatures of outflows (rather than inflows) in these sources.}
\label{v_med_figure}
\end{center}
\end{figure*}

We measure the velocity offset $\Delta $ between $v_{DAP}$ and $v_{med}$, where $v_{med}$ is the median velocity of the [OIII] profile from our multi-Gaussian fit and $v_{DAP}$ is the median velocity from the single-Gaussian DAP fit. For a pure Gaussian profile or a profile with an insignificant secondary component $v_{DAP} \simeq v_{med}$ and $\Delta v = 0$. In Figure \ref{v_med_figure} we show the distribution of $\Delta v$ values for all MaNGA spaxels that have a velocity width W$_{80}$ greater than the mean W$_{80}$ of their galaxy. This selection ensures that we only consider spaxels where the velocity width is indicative of out- or inflowing components. Figure \ref{v_med_figure} shows that the distribution of $\Delta v$ is slightly skewed towards negative $\Delta v$ values when considering all MaNGA galaxies or only non-AGN in MaNGA. 
However, the distribution of $\Delta v$ values for the MaNGA-selected AGN candidates is much more heavily skewed towards negative values. Although sometimes redshifted emission line components have been associated with (AGN-driven) outflows \citep[e.g.][]{Mueller_Sanchez_2011, Fischer_2013}, because of dust attenuation blue-shifted emission line profiles are a better probe and signature of outflowing gas.

We cannot fully exclude that in some galaxies the enhanced prevalence of secondary kinematic components in the [OIII] profile is not due to AGN-driven outflows, but rather due to signatures from mergers, bars or even inflows. But Figure \ref{v_med_figure} shows that most secondary kinematic components in MaNGA-selected AGN are blue-shifted secondary components to the [OIII] profile. This is an indication that we predominantly observe the signatures of outflows.

\subsection{Non-AGN galaxies with outflow signatures}

While we have shown that $\sim 2-3$ as many MaNGA-selected AGN show enhanced [OIII] kinematics compared to MaNGA galaxies not selected as AGN, a significant number of MaNGA non-AGN also show enhanced [OIII] kinematics. We visually inspect the spatially resolved MaNGA-BPT maps and the corresponding maps of the H$\alpha$ equivalent width (EW(H$\alpha$)) of all MaNGA non-AGN with $\langle$W$_{80}\rangle > 500$~km/s. All of those show BPT diagnostics consistent with AGN and/or LI(N)ER-like line ratios. However, they exhibit extremely low H$\alpha$ equivalent widths with EW(H$\alpha$)~$ < 3$~\AA. 

\citet{Cid-Fernandes_2010} have shown that invoking the H$\alpha$ equivalent width allows to differentiate between the different ionisation mechanisms that lead to the overlap in the LI(N)ER region of traditional diagnostic diagrams. Based on the bimodal distribution of EW(H$\alpha$), \citet{Cid-Fernandes_2010} suggest that EW(H$\alpha$)~$> 3$~\AA\ optimally separates true AGNs from `fake' AGNs in which LI(N)ER emission is due to hot evolved stars. As described in detail in \citet{wyle18}, the MaNGA AGN selection takes this additional criterion into account which is why such sources have not been selected as AGN candidates. We observe now that in addition to the true AGN-selected candidates these `fake' AGN make up the high velocity width tail from Figure \ref{W80_hist_mean}. This raises the question of what is driving these kinematic peculiarities. 

\begin{figure}
\begin{center}
\includegraphics[width = 0.5\textwidth, trim = 1cm 0.2cm 0cm 2cm, clip= true]{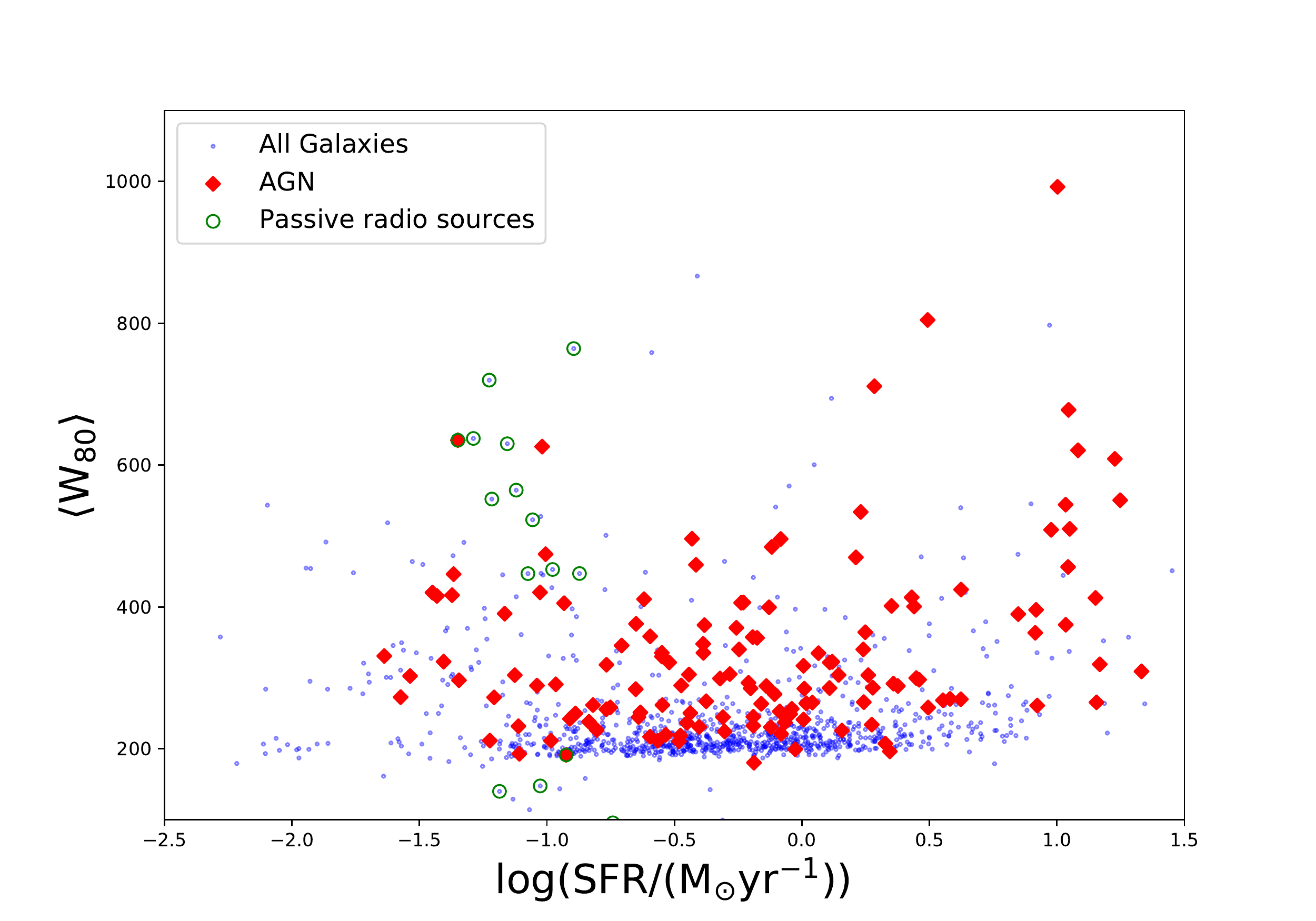}
\caption{Relation between the mean W$_{80}$ measurements $\langle$W$_{80}\rangle$ and the star formation rates in all MaNGA galaxies (blue cirlces) and in the MaNGA-selected AGN (red diamonds). No significant correlation is observed in either sample suggesting that it is not stellar-driven winds that are the dominant reason for high $\langle$W$_{80}\rangle$ in both the all MaNGa sample and the MaNGA-selected AGN. Only sources that pass the quality cut described in Section 4.1 are plotted here.}
\label{sfr_w80}
\end{center}
\end{figure}

Typical sources that have to be considered when observing [OIII] velocity widths exceeding $500$~km/s are mergers \citep{alag12, harr12} and outflows driven by stellar processes \citep{heck15}. Alternatively, these sources may host AGN that have not been identified as AGN in the optical selection \citep{wyle18} or relic AGN \citep{Ishibashi15} in which outflow signatures of a previous AGN episode are still imprinted on the gas kinematics. But a visual inspection of the optical images of these sources and the star formation rates (see Fig. \ref{sfr_w80}) do not suggest that mergers or stellar-driven outflows are a major contributor to the observed kinematic peculiarities.

In Figure \ref{sfr_w80} we furthermore mark objects that have been identified as `passive radio sources' (Wylezalek et al. in prep.). These sources have infrared Vega colors, measured with the \textit{Wide-field Infrared Survey Explorer (WISE)} satellite, of $-0.2 < \rm{W1-W2} < 0.3$ and $0 < \rm{W2-W3} < 2$, indicative of a passive galaxy population and no AGN activity \citep{ster12b}. But these objects have high radio-to-IR flux ratios. This type of objects, potentially low-luminosity radio AGN, are sometimes associated with late feedback from AGN required to be strong enough to suppress the late cooling flows of hot gas and keep quiescent galaxies red and dead \citep{heck14}. We now find that some of these objects make up the low SFR, high $\langle$W$_{80}\rangle$ tail in Figure \ref{sfr_w80}, suggesting that some of these identified high $\langle$W$_{80}\rangle$ sources are indeed active AGN missed in the optical selection. In Wylezalek et al. in prep., we explore in detail the relation between the multi-wavelength properties of AGN identified using various selection criteria and their gas kinematics. 

\subsection{AGN with no outflow signatures}

It is also notable, that many MaNGA-selected AGN candidates in this analysis do not show striking outflow signatures. \citet{Nedelchev17} investigated nearly 10000 Seyfert galaxies selected from SDSS DR7 to look for cold-gas outflows traced by Na D, finding outflow signatures in only 0.5\% of the population compared to 0.8\% of the control galaxies. They conclude that nearby optical AGN rarely drive kpc-scale cold-gas outflows and are not more frequent than in non-AGN. Similarly, \citet{Roberts-Borsani_2018} recently conducted a stacking analysis of the Na D doublet of 240,567 inactive galaxies and 67,753 AGN hosts from the SDSS DR7 survey to probe the prevalence and characteristics of cold galactic-scale flows local galaxies. They find little variation in either outflow or inflow detection rate between the AGN and non-AGN hosts. While this appears somewhat at odds with the findings in this paper, there are several plausible reasons for these different findings. 

The first and most obvious one is that we may not be observing the same type of AGN and not the same type of outflows. As described in detail in \citet{wyle18}, the AGN selection utilised in this paper is sensitive to a more nuanced picture of AGN activity and due to the IFU nature of the MaNGA survey, the selection allows to discover AGN signatures at large distances from the galaxy center. In particular, the selection allows to identify AGN candidates which have been missed in the previous SDSS single fibre surveys. Additionally, the analysis in this paper focuses on ionised gas signatures whereas both studies cited above investigate the prevalence of neutral outflow/inflow signatures in local galaxies and AGN. While both simulations and observations show that AGN-driven outflows are multiphase phenomena, the actual link between the different gas phases participating in the outflow remains unknown \citep{guil12, rupk13a, Costa_2015, Santoro_2018}, especially because the AGN may ionise significant fractions of the neutral gas reservoir. Therefore, single-phase studies often lead to wrong or incomplete estimates of the prevalence, extent, mass, and energetics of outflows and may therefore lead to misinterpreting their relevance in galaxy evolution \citep{Cicone_2018}.

Na~D absorption can, for example, be subject to significant uncertainties due to the fact that it can be probed only where there is enough background stellar light. Studying molecular outflows and their connection to ionised and neutral atomic phases in a sample of 45 local galaxies, \citet{Fluetsch_2018} compute the ratio of the molecular to atomic neutral outflow rates $\dot{M}_{H_{2}}/ \dot{M}_{HI}$. They use both the Na~D absorption to compute neutral mass outflows rates and the fine-structure line of C+, [CII]$\lambda$157.74$\mu$m, which is an alternative way to probe  atomic neutral outflows. They find $\dot{M}_{H_{2}}/ \dot{M}_{HI}$ to be an order of magnitude larger when using Na D absorption as tracer for $\dot{M}_{HI}$ compared to using [CII] as tracer for $\dot{M}_{HI}$. This comparison suggests that outflow detections and outflow mass measurements based on Na~D absorption are likely lower limits and that a multiwavelength and multi-phase assessment would lead to more complete and higher values/rates \citep{Roberts-Borsani_2018}. 

In this work, we do find a significant difference between the prevalence of ionised outflow signatures in MaNGA-selected AGN compared to non-AGN with larger prevalence fractions than reported in the above cited works. For example, we find that 25/7/2\% of MaNGA-selected AGN candidates have $\langle$W$_{80}\rangle > 500/800/1000$~km/s (see Section 4.1). However, the outflow prevalence fractions are still quite low. One of the reasons for this observation is certainly connected to the type of AGN we are probing, which are primarily weak AGN. Theoretical models \citep{zubo12} suggest that AGN need to provide sufficient luminosity to be able to push the gas out of the galactic potential. This `threshold' nature of AGN feedback was recently also suggested by molecular gas \citep{veil13a} and radio observations \citep{zaka14}. Additionally, our observations are only sensitive to outflows on galaxy-wide scales. Small-scale outflows with the size of a few kpc would be missed in this analysis due to the limitations in spatial resolution \citep{wyle17a}.

\begin{figure}
\begin{center}
\includegraphics[width = 0.5\textwidth, trim = 0cm 0.4cm 0cm 2cm, clip= true]{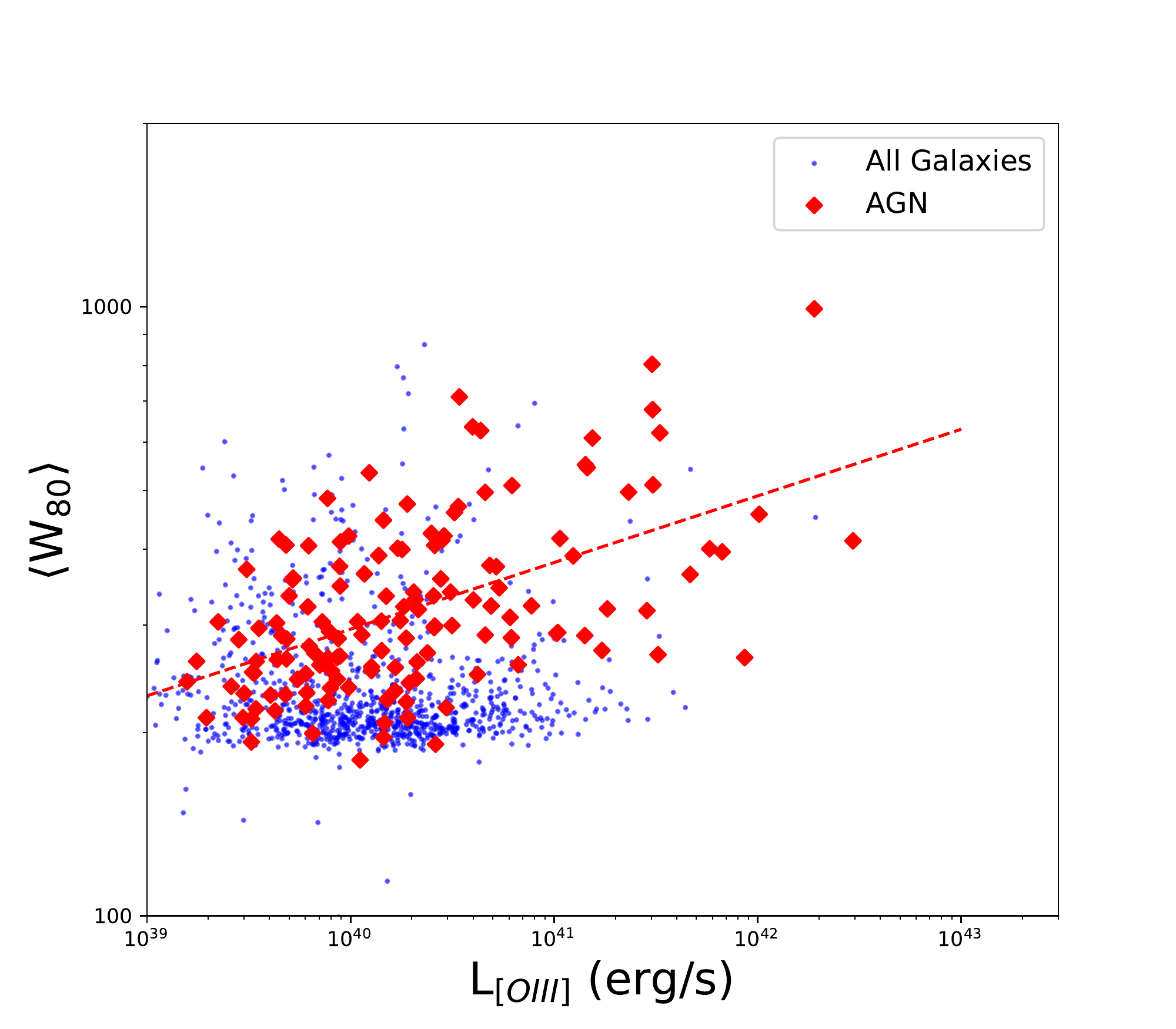}
\caption{Relation between the mean W$_{80}$ measurements $\langle$W$_{80}\rangle $and the total [OIII] luminosities $L_{[OIII]}$ (an indicator for the bolometric AGN luminosity) in all MaNGA galaxies (blue cirlces) and in the MaNGA-selected AGN (red diamonds). We observe a significant positive correlation between $\langle$W$_{80}\rangle$ and $L_{[OIII]}$ for the MaNGA-selected AGN suggesting that the AGN luminosity plays a dominant role in in the launching and detection of winds. Only sources that pass the quality cut described in Section 4.1 are plotted here.}
\label{oiii_w80}
\end{center}
\end{figure}

In Figure \ref{oiii_w80}, we show the relation between $\langle$W$_{80}\rangle$ and the total [OIII] luminosity L$_{[OIII]}$ as measured from the MaNGA observations. Although $L_{[OIII]}$ can be affected by extinction, most often from dust in the narrow-line region, [O III] luminosities are a good indicator of total bolometric AGN luminosity \citep{reye08, lama10}. We observe a significant positive correlation between $\langle$W$_{80}\rangle$ and $L_{[OIII]}$ for the MaNGA-selected AGN in this work. A Spearman rank test results in a $p-$value of $1.3 \times 10^{-10}$. We do not observe such a significant correlation for the whole MaNGA sample ($p-$value $=0.7$). This result confirms previous observations that AGN luminosity plays a dominant role in the launching and detection of outflows \citep{zaka14}. Furthermore, \citet{zubo18} suggests that in galaxies with low gas fraction, typically low redshift galaxies, AGN are fed by intermittent gas reservoirs, and thus the typical AGN episode duration is short \citep[for low-z AGN it is expected to be on the order of $10^5$ yrs, ][]{schawinski15, king15}. Since such host galaxies are mostly devoid of gas, any outflow inflated by the AGN is difficult to detect because they are faint. With the MaNGA survey primarily targeting low redshift, low- and intermediate luminosity AGN, it is not unexpected that only a small fraction of MaNGA-selected AGN selected exhibit clear and significant outflow signatures.

\subsection{Spatial distribution of the [OIII] velocity widths}

\begin{figure}
\begin{center}
\includegraphics[width=0.48\textwidth, trim = 1.8cm 2.2cm 2cm 2cm, clip= true]{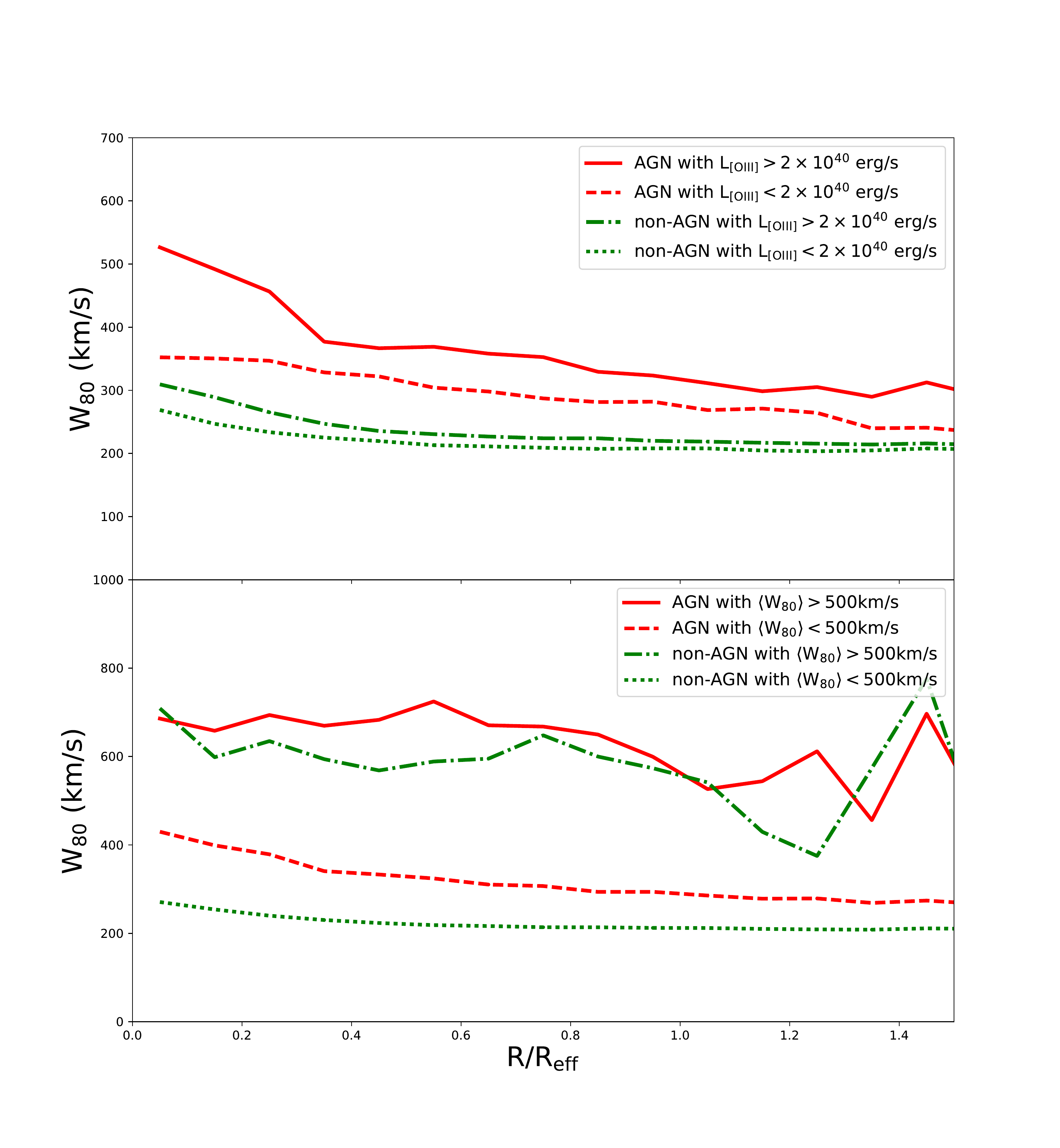}
\caption{Averaged radial dependence of W$_{80}$ as a function of projected distance from the galaxies' centres in units of effective radii R$_{\rm{eff}}$. \textbf{Top Panel}: Averaged W$_{80}$ radial profiles for the AGN and non-AGN samples, split by L$_{\rm{[OIII]}}$, respectively. \textbf{Bottom Panel}: Averaged W$_{80}$ radial profiles for the AGN and non-AGN samples, split by $\langle$W$_{80}\rangle$, respectively.}
\label{radial_profiles}
\end{center}
\end{figure}

We furthermore investigate the spatial distributions of the W$_{80}$ measurements. As above, we limit our analysis to MaNGA galaxies in which at least 10\% of the spaxels have valid [OIII] emission line measurements, i.e. a peak $S/N > 10$. We then measure the mean W$_{80}$ as a function of projected distance from the galaxies' centres in units of effective radii R$_{\rm{eff}}$. By design, all MaNGA galaxies are covered by the MaNGA footprint out to at least 1.5 R$_{\rm{eff}}$. In Figure \ref{radial_profiles}, we show the resulting radial profiles for the AGN and non-AGN samples, split by $\langle$W$_{80}\rangle$ and L$_{\rm{[OIII]}}$, respectively. 


	We observe a strong dependence of the AGN radial profiles on L$_{\rm{[OIII]}}$ where the AGN with L$_{\rm{[OIII]}} > 2 \times 10^{40}$~erg/s exhibit a steep rise in W$_{80}$ within the inner 0.4~R$_{\rm{eff}}$, indicative of enhanced [OIII] kinematics in the galaxy nuclei of luminous MaNGA AGN (Figure \ref{radial_profiles}, top panel). While the less luminous MaNGA AGN candidates do not show this steep rise in the center, both the low and higher luminosity AGN samples show higher W$_{80}$ measurements at all radii than the non-AGN samples. The non-AGN samples show mostly flat radial W$_{80}$ profiles. We observe a tentative rise at $\rm{R}< 0.4~\rm{R}_{\rm{eff}}$ in the high luminosity non-AGN sample with L$_{\rm{[OIII]}}~>~2~\times~10^{40}$~erg/s, showing that the non-AGN sample either includes AGN that have been missed in the optical selection (see Section 5.2) and/or sources with strong nuclear starbursts driving enhanced [OIII] kinematics.
	
	Separating the AGN and non-AGN samples by their overall mean [OIII] velocity width $\langle$W$_{80}\rangle$ (Figure \ref{radial_profiles}, bottom panel), we observe that the radial profiles of both the AGN and non-AGN samples with $\langle$W$_{80}\rangle > 500$~km/s are similar across all radii. Interestingly, neither of these profiles exhibits a significant rise towards the center. Rather, both profiles are relatively flat out to  $\rm{R} = 1.5~\rm{R}_{\rm{eff}}$. Visually investigating the [OIII] velocity width maps of these high $\langle$W$_{80}\rangle$ sources, we indeed confirm that the high velocity width signatures encompass the MaNGA footprint, indicative of large-scale enhanced [OIII] kinematics, possibly related to galaxy-wide outflows. In these cases the radial extent of the outflows likely exceeds the probed 1.5 R$_{\rm{eff}}$ such that the outflow size probed here can only be regarded as a lower limit. These are not necessarily the most L$_{\rm{[OIII]}}$ luminous sources as the difference of radial profiles to the AGN and non-AGN samples with  L$_{\rm{[OIII]}} > 2 \times 10^{40}$~erg/s shows. While we have shown in Figure \ref{oiii_w80} that L$_{\rm{[OIII]}}$ generally correlates with $\langle$W$_{80}\rangle$, consistent with what has been found in other works \cite[e.g.][]{zaka14}, there are some sources within both the AGN and non-AGN MaNGA samples with low- to intermediate [OIII] luminosities that show evidence of large-scale disturbed [OIII] kinematics. As discussed in Section 5.2 and shown in Figure \ref{sfr_w80}, the `passive radio sources', possibly missed low-luminosity radio AGN, make up a large fraction of the high $\langle$W$_{80}\rangle$ population within the non-AGN sample. 
	
	The radial velocity width profiles of the AGN and non-AGN samples with $\langle$W$_{80}\rangle < 500$~km/s show significantly lower W$_{80}$ measurements across all radii compared to the $\langle$W$_{80}\rangle > 500$~km/s samples. But the AGN sample with $\langle$W$_{80}\rangle < 500$~km/s consistently shows higher W$_{80}$ measurements compared to the non-AGN sample with $\langle$W$_{80}\rangle < 500$~km/s, with a slight rise of the radial profile at $\rm{R}< 0.4~\rm{R}_{\rm{eff}}$. This confirms our previous observations that the MaNGA-selected AGN in this work show distinct [OIII] kinematics from the non-AGN sample with a higher prevalence of enhanced [OIII] kinematics. 
	
	Several attempts have been made in the literature to quantify the size of AGN outflow regions and its relation to AGN power or galactic potential. However, this proves to be a difficult exercise as it is unclear what defines the `size' of an outflow. For example, outflow sizes can be defined as the spatial extent of a region above a certain velocity width threshold \citep{sun17} or edges of outflowing bubbles \citep[see][and references therein]{Harrison_2018}. Alternatively, outflow models assuming certain geometrical shapes of the outflowing region can be used to infer the maximum extent of the outflowing regions and deproject velocities. For example, \citet{wyle17a} has used a bi-cone model described in \citet{Mueller_Sanchez_2011} which consists of two symmetrical hollow cones having interior and exterior walls with apexes coincident with a central point source to model the outflow in an intermediate-luminosity MaNGA-selected AGN. However, this detailed modelling was only enabled by higher resolution follow-up IFU observations and is not available for all our source. 
	
\begin{figure}
\begin{center}
\includegraphics[width=0.48\textwidth, trim = 1.8cm 0.5cm 2cm 2cm, clip= true]{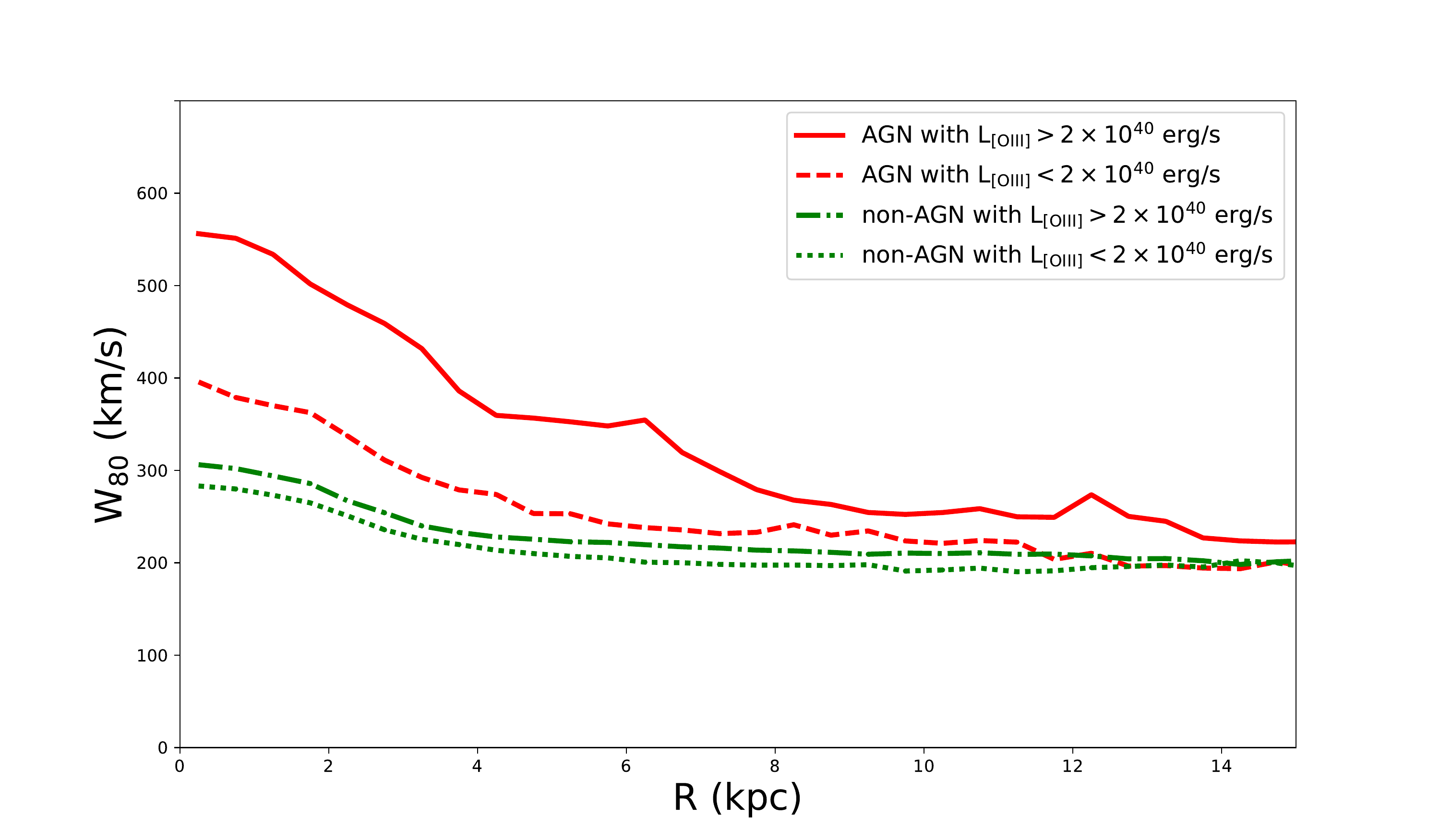}
\caption{Averaged radial dependence of W$_{80}$ as a function of projected distance from the galaxies' centres in units of kpc. We show the radial profiles for the AGN and non-AGN samples, split by L$_{\rm{[OIII]}}$, respectively.}
\label{radial_profiles_kpc}
\end{center}
\end{figure}

	Here, we quantify the size of the outflow signatures in AGN in the following way. In Figure \ref{radial_profiles_kpc} we show the averaged W$_{\rm{80}}$ radial profiles for the different L$_{\rm{[OIII]}}$ regimes as a function of absolute distance in kpc. Due to the different sizes of MaNGA galaxies, these radial profiles only provide a qualitative measurement of the difference in `sizes' of the outflow region. We observe that in AGN with L$_{\rm{[OIII]}} < 2 \times 10^{40}$~erg/s the averaged radial profile reaches the level of the non-AGN at $\sim 8$~kpc, whereas the radial profile of the AGN with L$_{\rm{[OIII]}} > 2 \times 10^{40}$~erg/s is starting to reach the level of the non-AGN samples at $\sim 15$~kpc. We do not show the radial profiles as a function of distance in kpc for the low and high $\langle$W$_{80}\rangle$ populations (as we did in Figure \ref{radial_profiles}) due to the small sample size of the high $\langle$W$_{80}\rangle$ population leading to noisy radial profiles beyond 8~kpc.
	
	The average [OIII] luminosities L$_{\rm{[OIII]}}$ of the two AGN samples shown in Figure \ref{radial_profiles_kpc} are $9\times10^{39}$~erg/s and $2\times10^{41}$~erg/s, respectively, while the average W$_{\rm{80}}$ measurements of the two AGN samples are 256~km/s and 335~km/s, respectively. For an outflow with a velocity $v_{gas}$ of these averaged W$_{\rm{80}}$ measurements, we can estimate the total kinetic energy injection rate over the lifetime $\tau$ of the ionised gas nebulae to be
	\begin{equation}
	\dot{E}_{kin} = \frac{E_{kin}}{\tau} =  \frac{M_{gas} v_{gas}^2}{2 \tau}   
	\end{equation}
	where $M_{gas}$ can be estimated from the [OIII] luminosity under the `Case B' assumption \citep{oste06, nesv11}, assuming that [OIII]/H$_{\beta} \sim 10$ \citep{liu13b} and adopting an electron density $n_e = 100~\rm{cm}^{-3}$\citep{gree11}. The gas mass can then be expressed as $M_{gas} = \left( \frac{L_{[OIII]}}{10^{44} \rm{erg~s^{-1}}}\right) \cdot 2.82 \times 10^9~\rm{M}_{\odot}$ \citep{liu13b}. The lifetime $\tau$ of the episode of AGN wind activity may be estimated as the travel time of clouds to reach the observed distances of 8/15~kpc from the centre traveling with the average outflow velocities $v_{gas}$ quoted above. These calculations yields total kinetic energy injection rates of $\dot{E}_{kin} \sim 2 \times 10^{38}$~erg/s and $\sim 5 \times 10 \times 10^{39}$~erg/s for the low and high luminosity AGN samples, respectively. 
	Although these calculations provide order-of-magnitude estimates at best, they show that \textit{(i)} even in low/intermediate luminosity AGN, the positive correlation between AGN power and outflow energetics persists as expected from models \citep{Dempsey_2018} and seen in observations \citep[][and references therein]{fior17} but that \textit{(ii)} the feedback processes in such low/intermediate luminosity, low redshift AGN probed by MaNGA are unlikely to have a significant impact on the evolution of their host galaxies, i.e. fully suppress star formation processes, as the kinetic coupling efficiencies $\dot{E}_{kin}/L_{AGN}$ are $\ll 1$\%  and most of them even likely $\ll 0.1$\%. Although some theoretical models suggest that efficiencies as low as 0.5\% may be sufficient in substantially suppressing star formation in the host \citep[e.g.][]{Hopkins_2010}, most models require efficiencies of $\sim 5$\% for feedback to be effective \citep[see][and references therein]{Harrison_2018}. However, our calculations are based on ionised gas observations which only probe the fraction of the gas that is in the warm ionised gas phase. This gas is likely in dense clouds that remain largely optically thick to the AGN ionising radiation and only a thin shell on the surface of these ionisation-bounded clouds produces emission lines \citep{Dempsey_2018}. Therefore, there may be additional outflow components that are not captured by our calculations. Over the lifetime of the AGN (typically $\sim 10^7$~yr) and depending on the amount of cool gas present in the individual host galaxies, such continuous and ubiquitous energy injection and outflow activity may still heat a fraction of the cool gas and delay or suppress star formation in individual cases \citep{Cheung_2016}.   
	
\subsection{Differences between AGN subsamples}

In Section 4.1, we analysed the three different AGN subsamples introduced. We have shown that the distributions of the mean W$_{80}$ measurements $\langle$W$_{80}\rangle$ of all three AGN subsamples are significantly different from the overall MaNGA distribution and the non-AGN distribution (see Figure \ref{W80_hist_mean} and Table \ref{table_statistics}), with the AGN distributions being more skewed towards higher $\langle$W$_{80}\rangle$ measurements. The same observation is true when comparing the distributions of the 75th percentile W$_{80}$ measurements W$_{80, 75th}$  (Figure \ref{W80_hist_75}). 

The Seyfert-AGN and SF-AGN are different from the full MaNGA and non-AGN distributions at a higher significance level (lower $p$-values) than the LINER-AGN distribution in both the $\langle$W$_{80}\rangle$ and the W$_{80, 75th}$ comparisons (Table \ref{table_statistics}). This observation, albeit tentative, might be explained by multiple factors. While the AGN selection method in \citep{wyle18} was developed such that `fake' AGN with LINER-like signatures would not be selected as AGN candidates, the authors noted that some of the LINER-AGN may be sources where the LINER-like photoionization signatures are connected to AGN-unrelated shocks. If indeed some of the selected LINER-AGN are related to shocks, while the shocks themselves may be due to AGN, stellar or merger activity \citep[see also][]{wyle17a}, then one would expect that the LINER-AGN sample is not as `clean' as the SF-AGN and Seyfert-AGN samples. Some of the LINER-AGN may in fact not be AGN. Therefore, LINER-AGN may show a lower rate of or different outflow signatures.

Generally, however, we observe little difference between the [OIII] kinematics of the three AGN subsamples. In particular, the Seyfert-AGN and SF-AGN subsamples show very similar characteristics in their [OIII] kinematics. The exact nature of the SF-AGN remains elusive. Various possibilities include recently turned off AGN, AGN dominated by a nuclear starburst, off-nuclear AGN and these are currently still being investigated (Wylezalek et al. in prep.). Our kinematic analysis suggests that the rate and nature of weak outflows is similar in both subsamples.

\section{Conclusion}

In this work we have examined the kinematics of the [OIII]$\lambda 5007 \AA$ emission line in each spatial element of 2778 low redshift galaxies observed as part of the SDSS-IV MaNGA survey. Specifically, we have developed a customised fitting method that allows to account for potential secondary kinematic components in the emission line profile as opposed to the MaNGA pipeline measurements that are based on single-Gaussian fitting. We first model the [OIII] emission line profile using both single and double-Gaussian profiles and evaluate the goodness of the fit based on its $\chi^{2}$ value. We then utilise the non-parametric measurements W$_{80}$ and $v_{med}$ to quantify the width of the emission line profile and the line-of-sight velocity. The main purpose of this work is to assess the prevalence of ionised gas outflow signatures in MaNGA-selected AGN candidates. Since MaNGA-observed AGN tend to be low- to intermediate luminosity AGN, a careful analysis of faint broad secondary components and/or deviations from a simple Gaussian profile need to be carefully assessed. 

Our strategy to do so is twofold. Only considering well detected emission lines with a signal-to-noise ratio $S/N > 10$, we first measure the mean $\langle$W$_{80}\rangle$ for each galaxy by averaging over all valid spaxels in that galaxy. We do not consider galaxies in which less than 10\% of the spaxels have valid [OIII] emission line measurements. We also determine the W$_{80}$ value that marks the 75th percentile of each galaxy's W80 distribution W$_{80, 75th}$. Both $\langle$W$_{80}\rangle$ and W$_{80, 75th}$ indicated in an absolute fashion which galaxies show large velocity widths. This type of analysis may miss galaxies with slow/moderate outflows that are imprinted as small deviations from a pure Gaussian [OIII] line profile. We therefore also assess the fractional change between the pipeline velocity width and the [OIII] velocity width derived in this work. 

Based on these derived quantities, we compare the [OIII] kinematics in the full MaNGA sample to the non-AGN in MaNGA, the MaNGA-selected AGN candidates and three subsamples of the MaNGA-selected AGN candidates. We find the following:

\begin{itemize}
\item The $\langle$W$_{80}\rangle$ and  W$_{80, 75th}$ distributions of the full MaNGA sample are significantly different from the MaNGA-selected AGN sample and from the individual AGN subsamples where MaNGA-selected AGN candidates tend to show higher $\langle$W$_{80}\rangle$ and W$_{80, 75th}$. 
\item Two to three times as many MaNGA-selected AGN candidates show enhanced [OIII] kinematics. This result is based on determining how many galaxies/AGN show W$_{80} > 500/800/1000$~km/s.
\item MaNGA-selected AGN require more often a multi-component model to describe their [OIII] emission line profiles compared to the non-AGN in MaNGA. While this result is more tentative, it suggests an enhancement of broad secondary components being present.
\item Comparing the line-of-sight velocities measured in this work to the line-of-sight velocities measured by the MaNGA pipeline, we find that the emission lines are predominantly blue-shifted suggesting that the kinematic peculiarities observed in this work are indeed related to outflows (rather than inflows). 
\item While generally AGN show a higher prevalence of ionised outflow signatures compared to the non-AGN in MaNGA, there are sources not selected as AGN that do show enhanced [OIII] kinematics. These sources do not display an enhanced merger fraction or indications that stellar processes might be driving these outflow indicators. Such sources may host AGN or AGN relics that have not been identified by the optical selection methods used here. A thorough multi-wavelength analysis is required to determine the cause of these enhanced [OIII] kinematics.
\item We observe a significant correlation between the [OIII] luminosity and  $\langle$W$_{80}\rangle$ in the MaNGA-selected AGN, confirming similar measurements in other works using other AGN samples. Based on these results it seems that AGN need to provide sufficient luminosity to be able to launch outflows and push the gas out of the galactic potential. Since most AGN in MaNGA are of low-/intermediate luminosity, it is therefore no surprise that we detect outflow signatures in only $\sim 25$\% of the MaNGA-selected AGN.
\item We find significant differences in the radial extent of broad [OIII] velocity components between the MaNGA-selected AGN and non-AGN sources. Higher luminosity AGN are able to drive larger scale outflows than lower luminosity AGN in agreement with in previous studies. The kinetic coupling efficiencies $\dot{E}_{kin}/L_{AGN}$ in MaNGA-selected AGN which are predominantly low- and intermediate luminosity sources are $\ll 1$\% which might imply that these AGN are unlikely to have a significant impact on the evolution of their host galaxies. However, these estimates are lower limits since we are likely missing a large fraction of the outflowing gas that is not in the here probed warm ionised gas phase. Over the lifetime of the AGN, continuous energy injection and outflow activity may still heat a fraction of the cool gas and delay or suppress star formation in individual cases even when the AGN is weak. 
\end{itemize}

This work shows that ionised outflow signatures are more prevalent in MaNGA-selected AGN than in non-AGN. Much of this work has only been possible due to the added spatial dimension provided by the MaNGA IFU data and shows that outflow and feedback signatures in low-luminosity, low-redshift AGN may previously have been underestimated.

\section*{Acknowledgements}

A.M.F. acknowledges the support of the NASA Maryland Space Grant Consortium. R.A.R. thanks partial financial support from CNPq and FAPERGS. 

Funding for the Sloan Digital Sky Survey IV has been provided by the Alfred P. Sloan Foundation, the U.S. Department of Energy Office of Science, and the Participating Institutions. SDSS-IV acknowledges
support and resources from the Center for High-Performance Computing at
the University of Utah. The SDSS web site is www.sdss.org.

SDSS-IV is managed by the Astrophysical Research Consortium for the 
Participating Institutions of the SDSS Collaboration including the 
Brazilian Participation Group, the Carnegie Institution for Science, 
Carnegie Mellon University, the Chilean Participation Group, the French Participation Group, Harvard-Smithsonian Center for Astrophysics, 
Instituto de Astrof\'isica de Canarias, The Johns Hopkins University, 
Kavli Institute for the Physics and Mathematics of the Universe (IPMU) / 
University of Tokyo, the Korean Participation Group, Lawrence Berkeley National Laboratory, 
Leibniz Institut f\"ur Astrophysik Potsdam (AIP),  
Max-Planck-Institut f\"ur Astronomie (MPIA Heidelberg), 
Max-Planck-Institut f\"ur Astrophysik (MPA Garching), 
Max-Planck-Institut f\"ur Extraterrestrische Physik (MPE), 
National Astronomical Observatories of China, New Mexico State University, 
New York University, University of Notre Dame, 
Observat\'ario Nacional / MCTI, The Ohio State University, 
Pennsylvania State University, Shanghai Astronomical Observatory, 
United Kingdom Participation Group,
Universidad Nacional Aut\'onoma de M\'exico, University of Arizona, 
University of Colorado Boulder, University of Oxford, University of Portsmouth, 
University of Utah, University of Virginia, University of Washington, University of Wisconsin, 
Vanderbilt University, and Yale University.

\bibliographystyle{mnras}

\bsp	
\label{lastpage}
\end{document}